\documentstyle[pra,aps,multicol,graphicx,subfigure]{revtex} 

\begin{document}

\title{Vortex Lattices in Stripe Domains  of  Ferromagnet/Superconductor 
Bilayer}

\author{Serkan Erdin\\
Department of Physics, Northern Illinois University, DeKalb, IL, 60115\\
\& Advanced Photon Source, Argonne National Laboratory,\\
 9700 South Cass Avenue,  Argonne, IL, 60439}

\maketitle

\begin{abstract}
The continuum theory of domain structures in ferromagnetic/superconducting 
bilayers fails when the equilibrium domain size becomes comparable with 
effective penetration depth $\Lambda$. Instead, a lattice of discrete 
vortices must be considered. Here, we report our results on the discrete 
vortex lattices in stripe domain structures of 
ferromagnetic/superconducting
bilayers. The vortices are assumed to be situated periodically on 
chains in stripe domains. We study the configurations 
containing up to 
three chains 
per domain,  and calculate their equilibrium energies, equilibrium domain 
size and vortex positions
through a method based on London-Maxwell equations. In equilibrium, the vortices
in the neighbor domains are half-way shifted while they are next to each other in the same domain.
Additionally, more vortex chains per domain appear spontaneously 
depending on  magnetization 
and domain wall energy. 
\end{abstract}

\vspace{1cm}

\noindent PACS Number(s): 74.25.Dw, 74.25.Ha, 74.25.Qt, 74.78.-w

\vspace{1cm}

\begin{multicols}{2}

\section{introduction}

Heterostructures made of superconducting (SC) and ferromagnetic (FM) 
pieces not only give rich physical effects that are not observed in 
individual subsystems, but also offer new devices that can be tuned by 
weak magnetic fields \cite{pok-rev,serk-rev}. One of the realizations of 
such 
heterostructures is ferromagnetic/suprconducting bilayer (FSB). 
Earlier Lyuksyutov and Pokrovsky noticed \cite{pok2,pok7} that in a
bilayer consisting of homogeneous SC and 
FM films with the magnetization normal to the plane, SC vortices occur
spontaneously in the ground state, even though the magnet does not
generate a magnetic field in the SC film. 

In previous work, we presented a
theory of such vortex-generation instability and the resulting vortex
structures \cite{stripe}. We showed that due to this instability, domains 
with
alternating magnetization and vortex directions occur in FSB.  
 In that study, we
treated these domain structures in the continuum regime in which the 
domain
size is much larger than the effective penetration depth, which is defined 
to be  $\Lambda = 
\lambda^2/d_{sc}$, where the London penetration depth $\lambda$ is much 
larger than the thickness of superconducting film $d_{sc}$ 
\cite{abrikosov}. Under the continuum aproximation, we found that the energy of 
stripe phase was minimal. The equilibrium domain size
and the equilibrium
energy for the stripe structure was found as \cite{stripe}
                                                                                                                                                             
\begin{equation}
L_{eq}^{(str)}=\frac{\Lambda }{4}\exp
\left( \frac{\varepsilon _{dw}}
{4\tilde{m}^{2}}-C+1\right),
\label{L-eq-stripe}
\end{equation}
\begin{equation}
{U_{eq}^{(str)}}
=-\frac{16\widetilde{m}^{2}
{\cal A}}{\Lambda }\exp
\left( -\frac{\varepsilon _{dw}}{4\widetilde{m}^{2}}
+C-1\right).
\label{U-eq-stripe}
\end{equation}
\noindent where $\tilde m = m -\varepsilon_v/\phi_0$,  $\varepsilon_v =
(\phi_0^2/16 \pi^2 \Lambda) \ln (\Lambda/\xi)$ is self-energy of a vortex,
$\varepsilon_{dw}$ is domain wall energy per domain wall length,$\cal A$
is the domain's area and $C
\sim 0.577$ is the
Euler-Mascheroni constant. If $\varepsilon_{dw}\leq 4\tilde{m}^2$, the continuum approximation
becomes invalid, since $L_{eq}$ becomes on the order of or less than
$\Lambda$ (see Eq.(\ref{L-eq-stripe})). However, it can be
recovered
by considering the discrete lattice of vortices instead.

In this paper, we study the discrete latice of vortices in stripe domain 
structures in FSB via a method based on London-Maxwell equations,
which is developed elsewhere \cite{serkan1}.
The  extension of the method to periodic systems is also introduced for 
the case of 
square magnetic
dot arrays on a SC film \cite{ser-physica}. Here we adapt it for the 
discrete 
vortex 
lattices in
SC/FM bilayers. In doing so, we assume that vortices and antivortices sit 
periodically on chains in the alternating domains of magnetization and 
vorticity. The problems we consider
here are; i) how the vortices and the antivortices are positioned on the
chains; ii) how the equilibrium domain size changes, depending on the
magnetization and the magnetic domain wall energy in the presence of the
vortices; iii) if the spontaneous appearence of domains structures with 
different number of vortex chains is possible. In order to solve these 
problems, we
first propose five different
configurations of the vortex and the antivortex chains, in which at most
two chains per stripe is considered.  Next, we calculate  their 
equilibrium
energies by means of numerical methods and find the most favorable case 
among them. 
 Our calculations show that
in equilibrium structure, vortex
chains are half
shifted in the adjacent domain while they are next to
each
other in the same  domains. Inspired from this result, similar
configuration for three vortex chains per domain is considered, and its
equilibrium energy is calculated for various values of  magnetization and domain
wall energy. Comparison of equilibrium energies of cases with one,two and
three chains per domains shows that at lower values of  magnetization and
domain wall energy case with two chains is favorable whereas , three
chain cases wins   at higher values of magnetization and domain wall 
energy.
Additionally, single chain case does not win over the ones with two and three 
chains per domain under any circumstances. 

The outline of this paper is as follows:
In the following section, we present the method for discrete case and its 
application to configurations with a single and double vortex chains per 
domain. In the third section, we briefly present our results on the 
proposed five  configurations. The fourth section is devoted to 
the study of 
 the case with three 
chains per domain.
The last section consists of the conclusions and discussion. In the 
appendix, 
we give the details of the methods and mathematical tricks in series 
calculations.

\section{Method}

In this section,
we introduce  a method based on the treatment of vortices in the discrete
lattice. We study the lattices of
discrete vortices only in the stripe phase.
In the continuum approximation, it is found that
the vortex density increases at the closer distances to the magnetic
domain walls. Based on this fact and the symmetry of the stripe domain
structure, it is reasonable to consider that the vortices and antivortices
form periodic structures on straight chains along the $y$ direction. Even
though it is not clear how many chains are associated with each domain, we
can still make progress
toward understanding discrete vortex lattices.
To this end, five stripe domain configurations
in which vortices are situated periodically on chains
 are proposed. From this point on, the configurations with $N$ vortex chains per 
stripe domain are labeled as $N$ state.

\begin{figure}
\centering
\subfigure[1st Case] 
{
    \label{fig:1stcase}
    \includegraphics[angle=0,width=3cm]{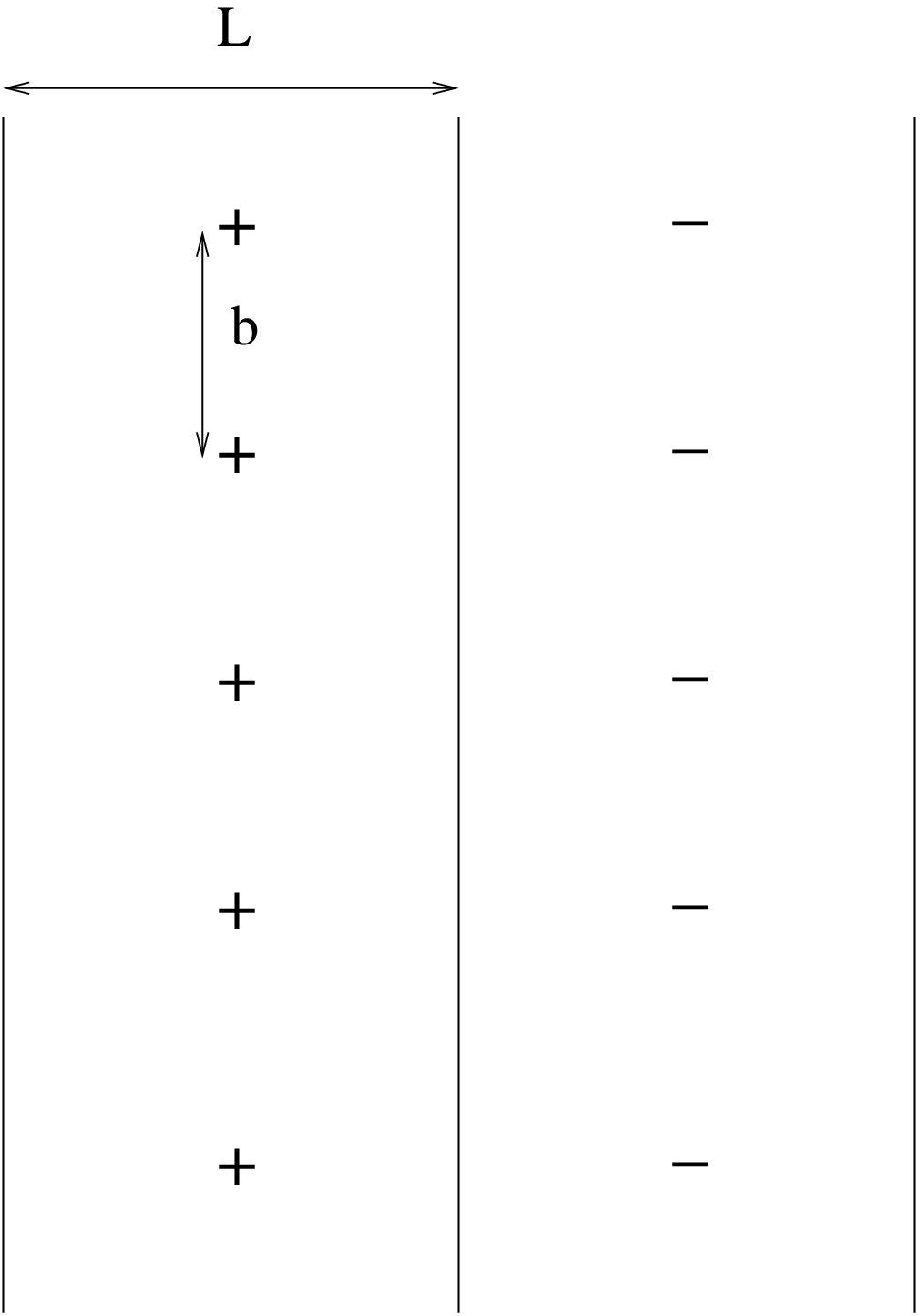}
}
\hspace{1cm}
\subfigure[2nd Case] 
{
    \label{fig:2ndcase}
    \includegraphics[angle=0,width=3cm]{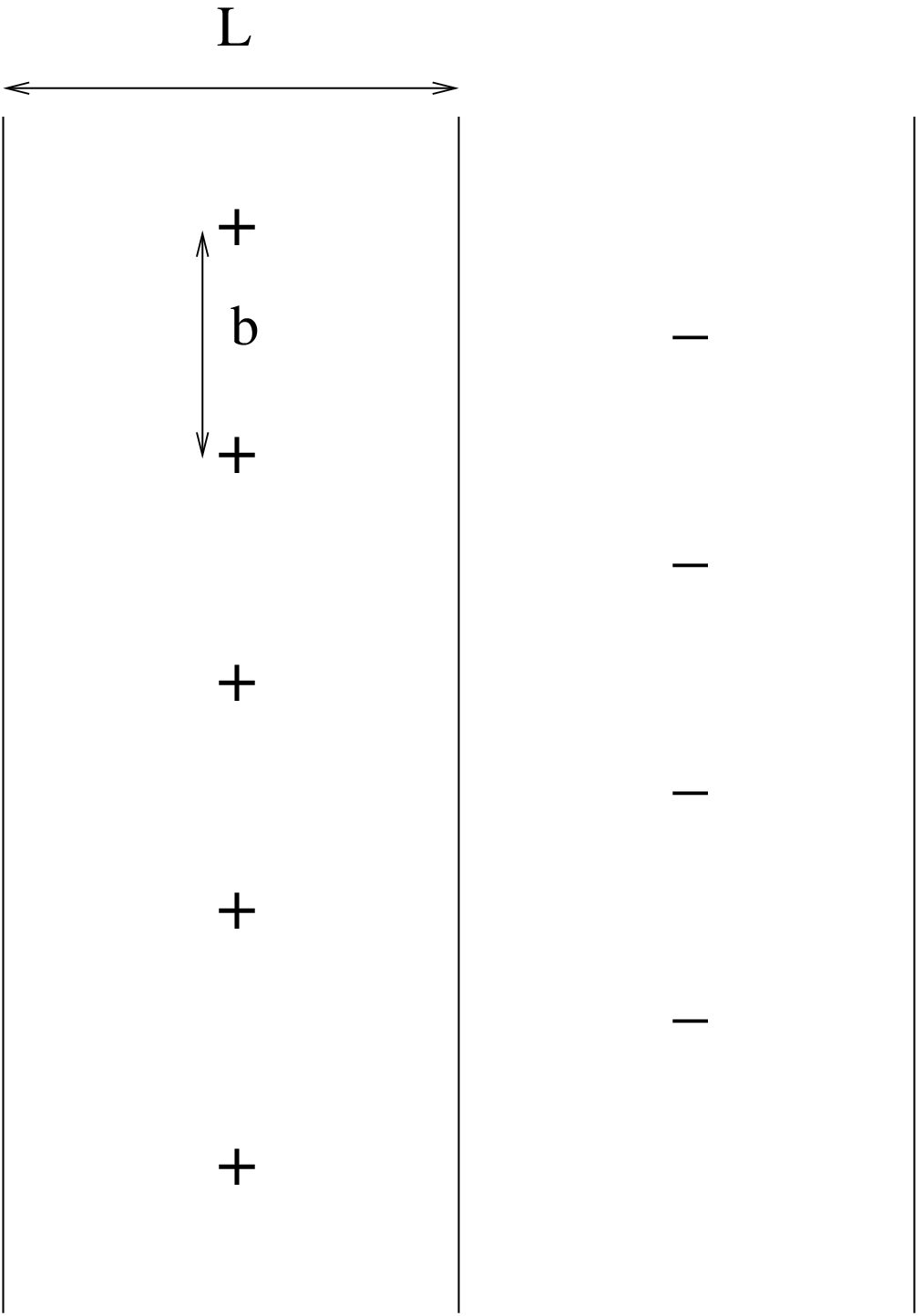}
}
\vspace{1cm}
\subfigure[3rd Case:1] 
{
    \label{fig:3rdcasea}
    \includegraphics[angle=0,width=3cm]{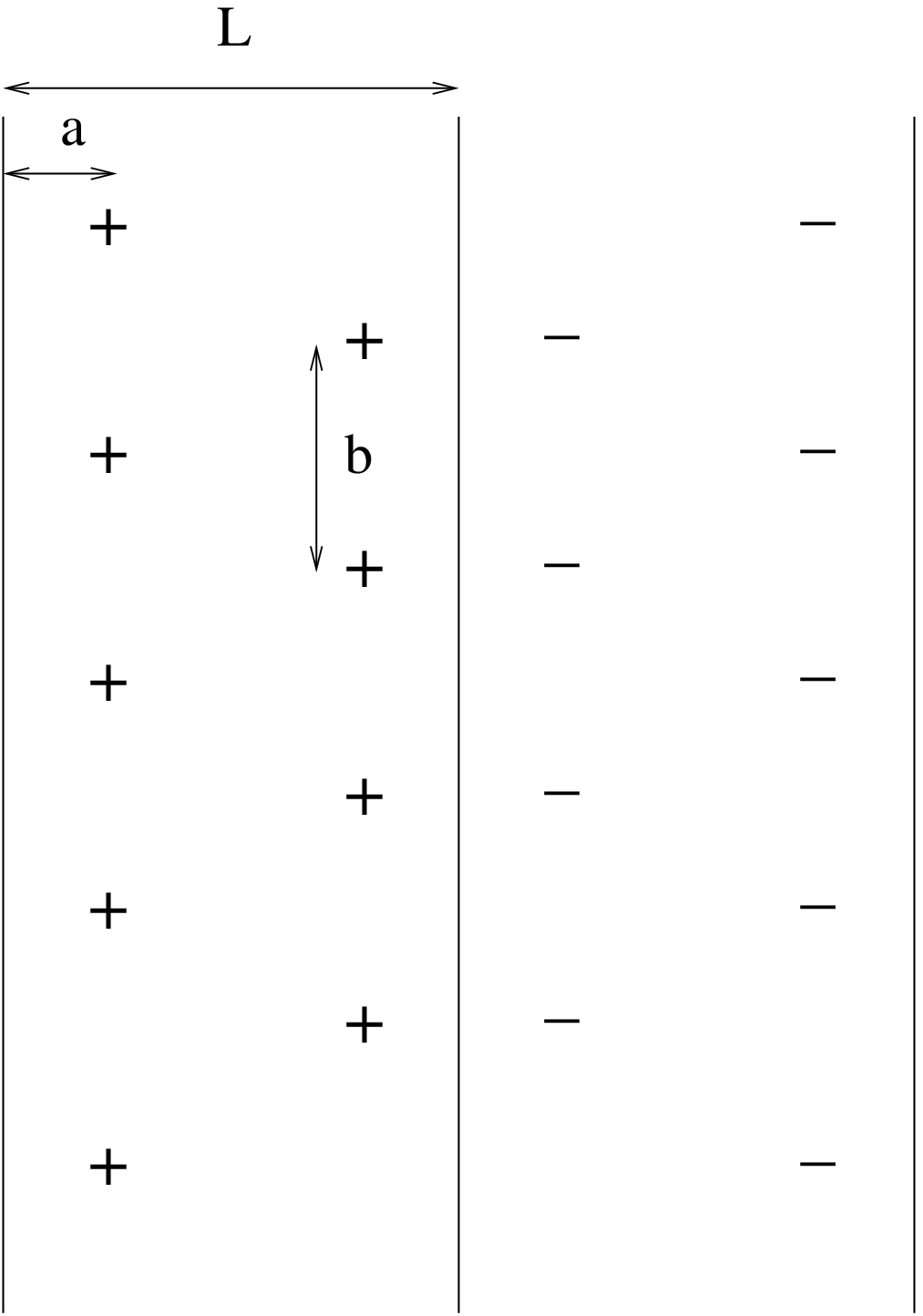}
}
\hspace{1cm}
\subfigure[3rd Case:2] 
{
    \label{fig:3rdcaseb}
    \includegraphics[angle=0,width=3cm]{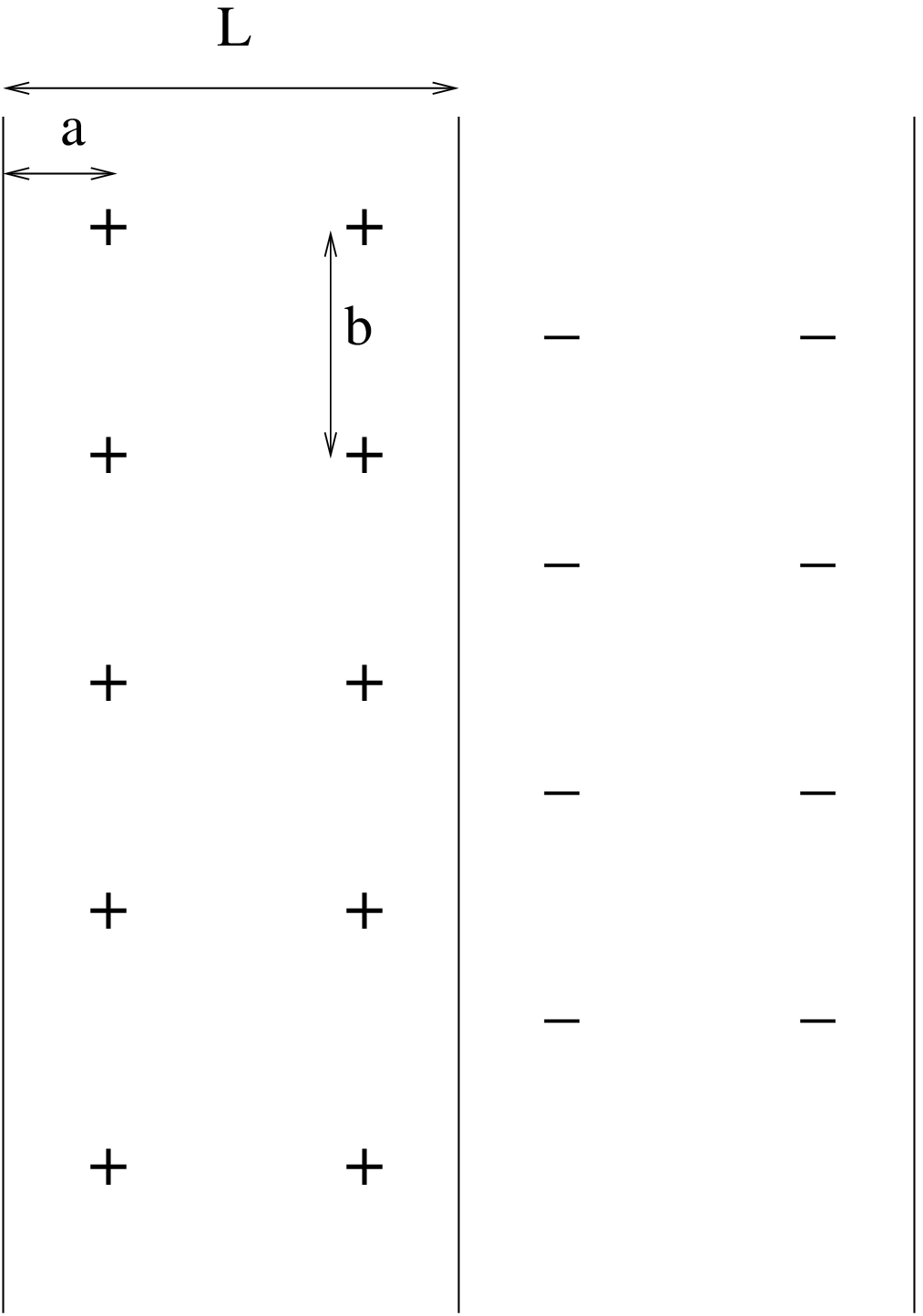}
}
\vspace{1cm}
\subfigure[4th Case] 
{
    \label{fig:4thcase}
    \includegraphics[angle=0,width=3cm]{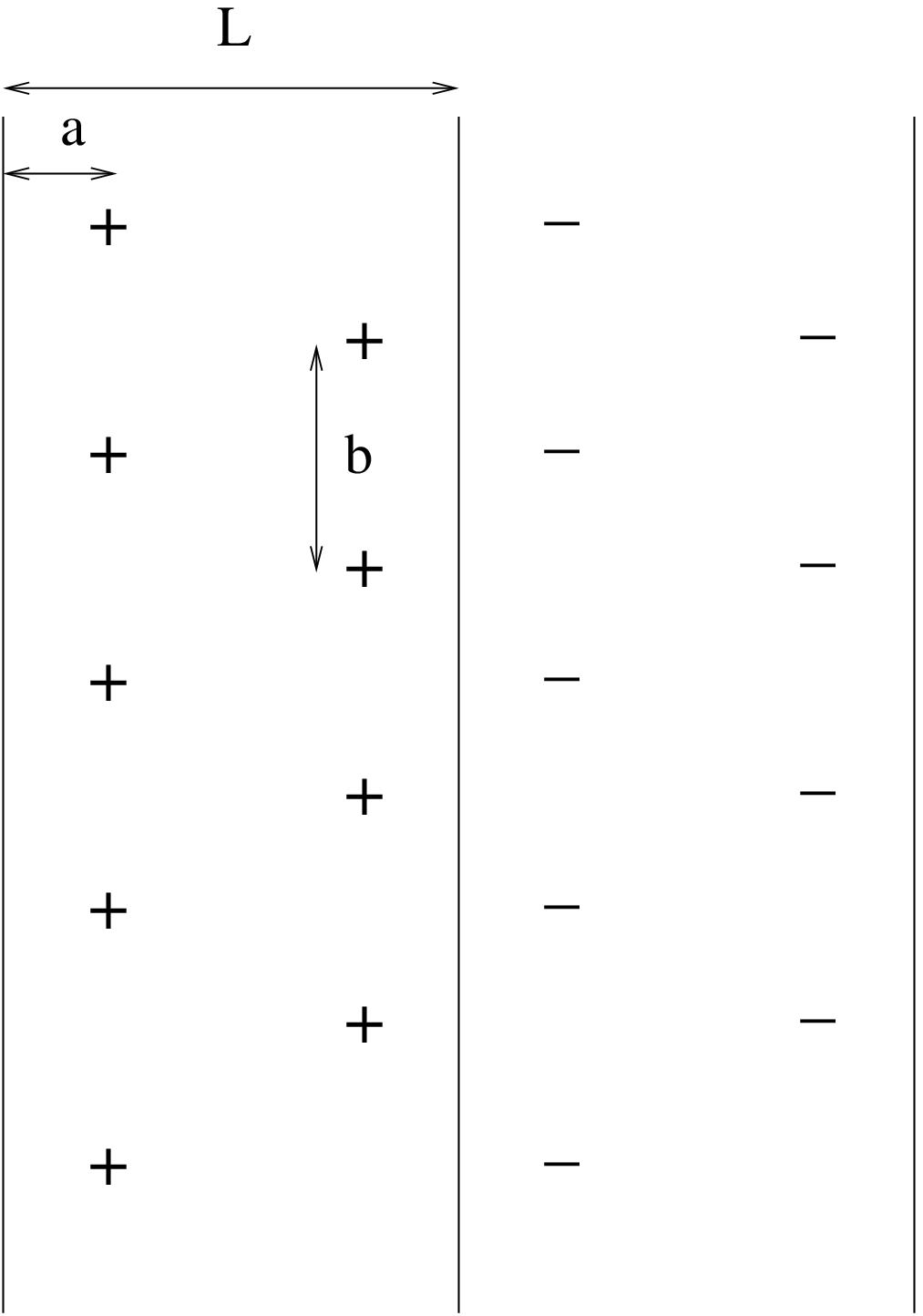}
}
\hspace{1cm}
\subfigure[5th Case] 
{
    \label{fig:5thcase}
    \includegraphics[angle=0,width=3cm]{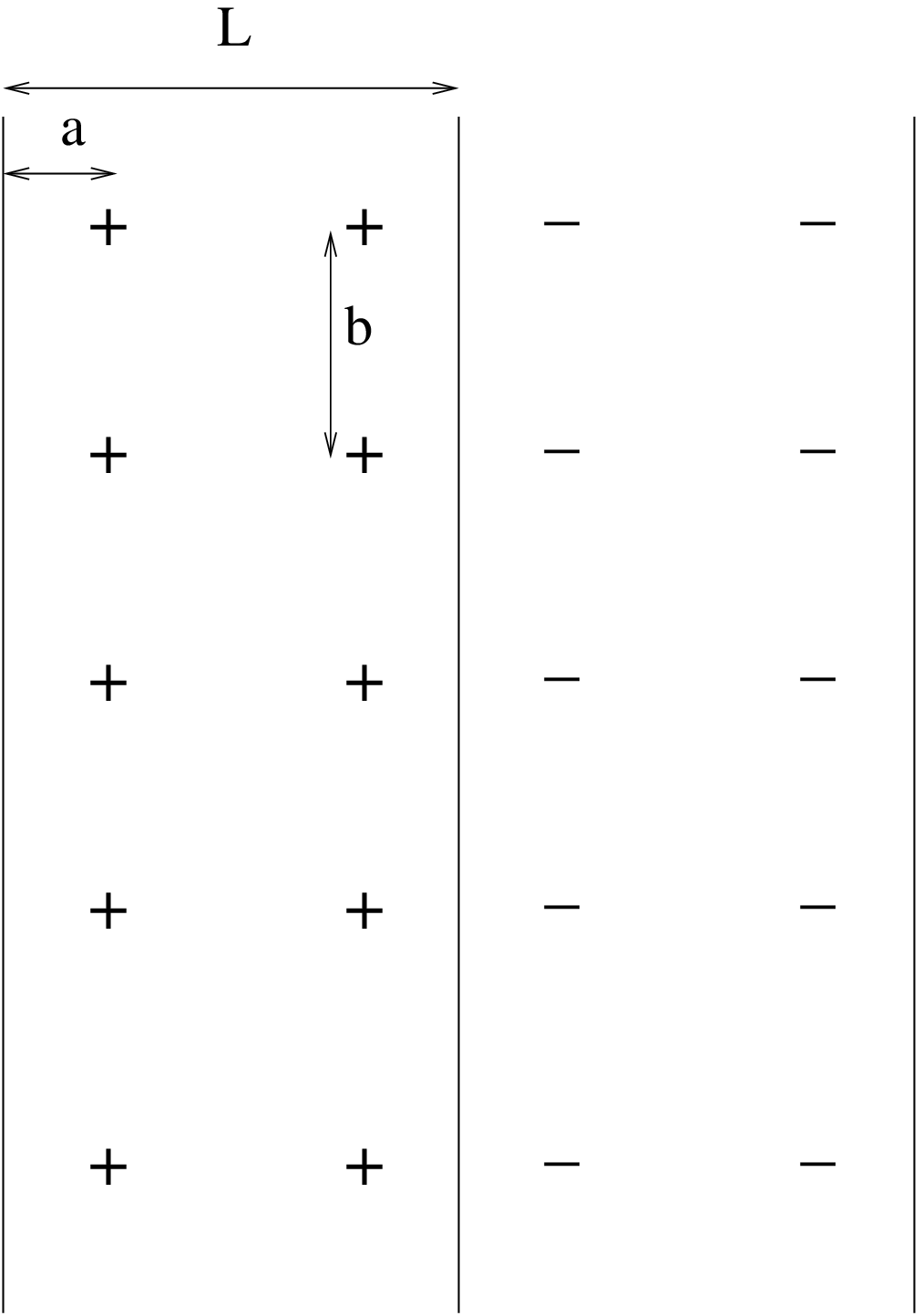}
}
\caption{Proposed configurations for $N=1$ and $N=2$ states.}
\label{fig:anal} 
\end{figure}

In two of proposed cases, there is one chain per stripe ($N=1$ states), located in the
middle of the domain. In this case, two configurations of vortex lattice
are possible; first, the vortices and the
antivortices in a neighboring domains are alongside to one another (see
Fig. \ref{fig:1stcase}), second, they are shifted by half period $b/2$ 
along the
$y$ direction, where $b$ is the distance between two nearest vortices on
the chain (see
Fig. \ref{fig:2ndcase}).


In the other three candidate lattice structures, there are two chains per
stripe domain $N=2$ states, at the distance $a$ from the magnetic domain walls. The
possible situations in the two chain cases are as follows; First, chains
in the same domains are shifted by a half period $b/2$ along $y$
direction (see
Fig. \ref{fig:3rdcasea})
. However, the neighbor vortices and antivortices are next to
each other. Second, they are just shifted by a half period $b/2$ 
(see
Fig. \ref{fig:4thcase}). Third, the vortices and the antivortices are 
simply side by
side on the chains (see
Fig. \ref{fig:5thcase}). 


Our next step is to write the energies of these five candidates. To this
end, we use the energy  equations for periodic systems developed elsewhere 
\cite{ser-physica}.   


\begin{eqnarray}
u_{vv} &=& \frac{\phi_0^2}{4 \pi{\cal A}^2} \sum_{\bf G}
\frac{ |F_{\bf G}|^2}{G ( 1 + 2 \lambda G)}, \label{puv} \\
u_{mv} &=& -
\frac{\phi_0}{\cal A} \sum_{\bf G} \frac{m_{z {\bf G}}F_{-\bf G}}{1 + 2
\lambda G}, \label{pumv}\\
u_{mm} &=& -2 \pi \lambda \sum_{\bf G} \frac{G^2
|{\bf m}_{z\bf G}|^2}{1 + 2 \lambda G}. \label{pumm}
\end{eqnarray}

In those equations, the vortex
configurations differ by their form-factors. We can obtain them from
$F_{\bf G} = \sum_{{\bf r}_i} n_i e^{i {\bf G} \cdot {\bf r}_i}$, where
the ${\bf G}$'s are the reciprocal vectors of the periodic structures, the
${\bf r}_i$ are the positions of the vortex centers, and $n_i$ are the
charge of the vortex. In our proposed models, ${\bf G} = ( ( 2 r + 1 )
\frac{\pi}{L}, 2 s \frac{\pi}{L})$ and $n_i = \pm 1$. Table \ref{tabF}
gives the form factors of each configuration in the order they are
described above.

\begin{table}[h] 
\caption{The form-factors of vortices in the proposed
configurations. \label{tabF}} 
\begin{center} 
\begin{tabular}{|c|c|}\hline
{ configuration}&{$F_{\bf G}$} \\ \hline 
1 & $i (-1)^r$ \\ \hline 
2 & $i (-1)^r ( 1 + (-1)^s)$ \\ \hline 
3 & $2 i \sin G_x a ( 1 + (-1)^s)$ \\ \hline 
4 & $e^{i G_x a} - (-1)^ s e^{- i G_x a}$ \\ \hline 
5 & $ 2 i \sin G_x a$ \\ \hline 
\end{tabular} 
\end{center} 
\end{table}

\noindent Note that, in Table \ref{tabF}, the form-factor for the third
configuration also belongs to the case in which the vortex and the
antivortex chains are shifted by half period only in the neighbour
domains, not in the same domain (see
Fig. \ref{fig:3rdcaseb}). Since information about the vortex
lattice is carried only by the form factors, there is no need to consider
the above-mentioned case separately.

In our calculations, the divergent part of the series must be extracted
carefully. We show below the detailed analysis of series equations for
each candidate. We start with the self interaction energy of the magnetic
layer $U_{mm}$, since it is the same for each configuration. For the
periodic structures, it is given by (\ref{pumm}). Direct substitution of
the Fourier coefficient of the stripe phase $m_{z{\bf G}}=\frac{2 i m
}{\pi( 2 r + 1)}$ into Eq.(\ref{pumm}) gives the self-interaction of the
magnetic layer per unit cell as

\begin{equation} 
u_{mm} = - \frac{8 m^2}{L} \sum_{r=0}^{\infty}
\frac{1}{\frac{L}{2 \pi \Lambda} + 2 r + 1}. \label{Hmmser} 
\end{equation}

\noindent where $\psi^{(0)} (x)$ is the polygamma function of zeroth order
\cite{abrom}.  In our numerical calculations, we write the logarithmic
term in (\ref{Hmmser}) as $\ln (\Lambda/l) + \ln (L/\Lambda)$ and then
incorporate the $-4m^2 \ln (\Lambda/l)$ term in the renormalized
$\varepsilon_{dw}^{ren}$. Another energy term with a divergent series is
the vortex energy, in general given by (\ref{puv}). The logarithmic
divergence in this term stems from the vortex self-energies. We first
split (\ref{puv})  into two parts as follows:

\begin{equation} 
u_{vv} = \frac{\pi \varepsilon_0}{2 L^2 b^2} \sum_{\bf
G}\left[ \frac{ |F_{\bf G}|^2}{G^2} - \frac{ |F_{\bf G}|^2}{G^2 (1 + 2
\Lambda G)}\right]. \label{vorener} 
\end{equation}

\noindent Note that the area of the unit cell is $2 L b$. The first
term of the series above contributes to the self-energies of the vortices;
whereas, the second term is the vortex-vortex energy and will be left in
the series form. For each form-factor in Table \ref{tabF} the series in
the first term can be transformed to the form of \\
$\sum_{r=-\infty}^{\infty}\sum_{s=-\infty}^{\infty} 1/((2 r + 1)^2 x^2 +
s^2)$ where $x$ is constant, and depends on the form-factor. A detailed
analysis of such series is given in Appendix A.

The next step is to find the vortex energy and the interaction energy of
the magnetization and vortices
 for each configuration. In the calculation of $u_{mv}$, we take the
Fourier coefficient of the magnetization to be $\frac{4 i m }{( 2 r +
1)}\delta (G_y)$. The fact that the stripe is infinite along the $y$
direction results in the additional term $2 \pi \delta (G_y)$. However, it
does not play any role in the calculation of $u_{mm}$. For numerical
analysis, these energies must be expressed in terms of dimensionless
parameters. To this end, we define dimensionless variables $\tilde \Lambda
= \Lambda/L$ , $\tilde b = b/L$ and $\tilde \varepsilon_{dw}
 = \varepsilon_{dw}^{ren} \Lambda/\varepsilon_0$.  The total energy
$\tilde U$ is measured in units of $\varepsilon_0/\Lambda^2$. In addition,
we introduce the dimensionless magnetic energy as $\tilde U_{mm} =
u_{mm}/(\varepsilon_0/\Lambda^2)$. In terms of these parameters, the
energy of the first configuration reads

\end{multicols}

\begin{equation} 
\tilde U^{(1)} = \frac{\tilde \Lambda^2}{4 \tilde b}
\left(\ln \left( \frac{4 \Lambda}{ e^C \tilde \Lambda \xi}\right) + 2 
f_v^{(1)} 
( \tilde
\Lambda) - \frac{2 f_{vv}^{(1)} ( \tilde \Lambda, \tilde b )}{ \tilde b
\pi} - \frac{16 m \phi_0}{\varepsilon_0} f_{mv}^{(1)} (\tilde \Lambda )
\right) + \tilde U_{mm} + \tilde \varepsilon_{dw} \tilde \Lambda,
\label{conf1} 
\end{equation}


\noindent where,

\begin{eqnarray} 
f_v^{(1)} &=& \sum_{r=0}^{\infty} \frac{\coth( ( 2 r + 1)
\frac{\tilde b \pi}{2}) -1}{2 r + 1}, \nonumber \\ 
f_{vv}^{(1)} &=&
\sum_{r,s = - \infty}^{\infty} \frac{1}{\left(( 2 r +1 )^2 + \frac{4
s^2}{\tilde b^2})(1+2 \pi \tilde \Lambda \sqrt{( 2 r +1 )^2 + \frac{4
s^2}{\tilde b^2}})\right)}, \nonumber \\
f_{mv}^{(1)} &=& \sum_{r=0}^{\infty} \frac{(-1)^r}{(2 r + 1 ) ( 1 + 2 \pi
\tilde \Lambda( 2 r + 1 ))}. \label{fmmfmv} 
\end{eqnarray}

\begin{multicols}{2}

The form factor for the second configuration survives only for even values
of $s$. Then, the dimensionless energy of the second configuration is
found to be

\end{multicols}

\begin{equation} 
\tilde U^{(2)} = \frac{\tilde \Lambda^2}{2\tilde b}
\left(\ln \left( \frac{4 \Lambda}{e^C \tilde \Lambda \xi}\right) + 
2f_v^{(2)} ( \tilde
\Lambda) - \frac{4 f_{vv}^{(2)} ( \tilde \Lambda, \tilde b ) }{\tilde b
\pi} - \frac{16 m \phi_0}{\varepsilon_0} f_{mv}^{(2)} (\tilde \Lambda
)\right) + \tilde U_{mm} + \tilde \varepsilon_{dw} \tilde \Lambda,
\label{conf2} 
\end{equation} 
\noindent where $f_{mv}^{(2)} = f_{mv}^{(1)}$ and,


\begin{eqnarray} 
f_v^{(2)} &=& \sum_{r=0}^{\infty} \frac{\coth (( 2 r + 1
)\frac{\tilde b\pi }{4}) -1}{2 r + 1}, \nonumber \\
f_{vv}^{(2)} &=& \sum_{r,s = - \infty}^{\infty} \frac{1}{\left(( 2 r +1 )^2 +
\frac{16 s^2}{\tilde b^2})(1+2 \pi \tilde \Lambda \sqrt{( 2 r +1 )^2 +
\frac{16 s^2}{\tilde b^2}})\right)}. \label{fvfmv2} 
\end{eqnarray}

\begin{multicols}{2}

In the third configuration, as in the second configuration, only even
values of $s$ contribute to the energy. In the first two configurations,
the square of their form-factors enter the vortex energy as a constant.
However, in this case, the square of the sine function appears. In
Appendix A, it is shown how to calculate the series in the presence of
such functions. Introducing the dimensionless parameter $\tilde a = a/L$,
the energy functional of the third configuration becomes

\end{multicols}

\begin{eqnarray} 
\tilde U^{(3)} &=& \frac{\tilde \Lambda^2}{\tilde b}
\left( \ln \left( \frac{4 \Lambda}{e^C \tilde \Lambda \xi}\right)-\ln 
(\cot(\pi \tilde a))
+ 4 f_v^{(3)} ( \tilde \Lambda, \tilde a) - \frac{8 }{\tilde b \pi}
f_{vv}^{(3)} ( \tilde \Lambda, \tilde a, \tilde b ) - \frac{16m
\phi_0}{\varepsilon_0} f_{mv}^{(3)} (\tilde \Lambda,\tilde a )\right) 
\nonumber \\
&+& \tilde U_{mm} + \tilde \varepsilon_{dw} \tilde \Lambda, \label{conf3}
\end{eqnarray}


\noindent where,

\begin{eqnarray} 
f_v^{(3)} &=& \sum_{r=0}^{\infty} \frac{\coth (( 2 r + 1
)\frac{\tilde b \pi}{4} ) -1}{2 r + 1} \sin^2 (( 2 r + 1 ) \pi \tilde a),
\nonumber \\
f_{vv}^{(3)} &=& \sum_{r,s = - \infty}^{\infty} \frac{\sin^2 (( 2 r + 1 )
\pi \tilde a) }{\left(( 2 r +1 )^2 + \frac{ 16 s^2}{\tilde b^2})(1+ 2 \pi
\tilde \Lambda \sqrt{( 2 r +1 )^2 + \frac{16 s^2}{\tilde b^2}})\right)},
\nonumber \\
f_{mv}^{(3)} &=& \sum_{r=0}^{\infty} \frac{\sin (( 2 r + 1 ) \pi \tilde
a)}{(2 r + 1 ) (1+2 \pi \tilde \Lambda ( 2 r + 1 ))}. \label{fvfmv3}
\end{eqnarray}

\begin{multicols}{2}

 In the fourth configuration, the square of the form-factor is: \\
$|F_{\bf G}|^2 = 2 - 2 (-1)^s \cos ( ( 2 r + 1 ) \pi \tilde a )$. Even and
odd values of $s$ give different contributions. Then, we can calculate the
vortex energy for even $s$ and odd $s$ separately. Employing similar
techniques, we find
 
\end{multicols}

\begin{equation} 
\tilde U^{(4)} = \frac{\tilde \Lambda^2}{2 \tilde b}
\left( \ln \left( \frac{4 \Lambda}{e^C \tilde \Lambda \xi}\right) + 2 
f_v^{(4)} ( \tilde
\Lambda, \tilde a) - \frac{4}{\tilde b \pi} f_{vv}^{(4)} ( \tilde \Lambda,
\tilde a, \tilde b ) - \frac{16m \phi_0}{\varepsilon_0} f_{mv}^{(4)}
(\tilde \Lambda,\tilde a )\right) + \tilde U_{mm} + \tilde 
\varepsilon_{dw} \tilde
\Lambda, \label{conf4} 
\end{equation}


\noindent where $f_{mv}^{(4)} = f_{mv}^{(3)}$ and,

\begin{eqnarray} 
f_v^{(4)} &=& \sum_{r=0}^{\infty} \frac{\coth (( 2 r + 1
)\frac{\pi \tilde b}{4} ) -1}{2 r + 1} \sin^2 (( 2 r + 1 ) \pi \tilde a)
+ \sum_{r=0}^{\infty} \frac{\tanh (( 2 r + 1 )\frac{\pi \tilde b}{4} )
-1}{2 r + 1} \cos^2 (( 2 r + 1 ) \pi \tilde a), \nonumber \\ 
f_{vv}^{(4)}
&=& \sum_{r,s = - \infty}^{\infty} \frac{\sin^2 (( 2 r + 1 ) \pi \tilde
a)}{\left(( 2 r +1 )^2 + \frac{16 s^2}{\tilde b^2})(1+ 2 \pi \tilde 
\Lambda
\sqrt{( 2 r +1 )^2 + \frac{16 s^2 }{\tilde b^2}})\right)} \nonumber \\
 &+& \sum_{r,s = - \infty}^{\infty}\frac{\cos^2 (( 2 r + 1 ) \pi \tilde
a)}{ \left(( 2 r +1 )^2 + \frac{4 ( 2 s+1 )^2}{\tilde b^2}) (1+ 2 \pi 
\tilde \Lambda
\sqrt{( 2 r +1 )^2 + \frac{4 (2 s +1)^2 }{\tilde b^2}})\right)}. 
\end{eqnarray}

\begin{multicols}{2}

The form-factor for the fifth case resembles that of the third case with
an exception. That is, in the third case, only even values of $s$ are
taken into account, while all integers contributes to the sum over $s$ in
the fifth case. Keeping this in mind, it is straightforward to
obtain the dimensionless energy for the last case as

\end{multicols}

\begin{eqnarray} 
\tilde U^{(5)} &=& \frac{\tilde \Lambda^2}{2 \tilde b}
\left( \ln \left( \frac{4 \Lambda}{e^C \tilde \Lambda \xi} \right) -\ln ( 
\cot(\pi \tilde
a) + 4 f_v^{(5)} ( \tilde \Lambda, \tilde a) - \frac{4}{\tilde b \pi}
f_{vv}^{(5)} ( \tilde \Lambda, \tilde a, \tilde b ) - \frac{16 m
\phi_0}{\varepsilon_0} f_{mv}^{(5)} (\tilde \Lambda ,\tilde a)\right) 
\nonumber \\
&+& \tilde U_{mm} + \tilde \varepsilon_{dw} \tilde \Lambda, \label{conf5}
\end{eqnarray}


\noindent where $f_{mv}^{(5)} = f_{mv}^{(3)}$ and, 
\begin{eqnarray}
f_v^{(5)} &=& \sum_{r=0}^{\infty} \frac{\coth (( 2 r + 1 )\frac{\pi \tilde
b}{2} ) -1}{2 r + 1} \sin^2 (( 2 r + 1 ) \pi \tilde a), \nonumber \\
 f_{vv}^{(5)} &=& \sum_{r,s = - \infty}^{\infty} \frac{ \sin^2 (( 2 r + 1
) \pi \tilde a)}{\left(( 2 r +1 )^2 + \frac{4 s^2}{\tilde b^2})(1+ 2 \pi \tilde
\Lambda \sqrt{( 2 r +1 )^2 + \frac{4 s^2}{\tilde b^2}})\right)}. 
\label{fvfmv5}
\end{eqnarray}

\begin{multicols}{2}

\section{Results for $N=1$ and $N=2$ States}

In this section, we present our results based on numerical calculations. 
The series in 
$f_v^{(i)}$ converges very fast when $r_{max} > 200$
, while the series in 
$f_{vv}^{(i)}$ and converges rather slowly, although,  when  
$r_{max}>4000$ and
$s_{max} >4000$, the results do not change up to 6th decimal point in the 
energy,  where $i$ labels the particular domain configuration.
To make sure of this accuracy in the calculations, 
we take $r_{max}= 600$ for $f_v^{(i)}$,  $r_{max}= 5000$
for $f_{mv}^{(i)}$,  and $r_{max}= 5000$ and
$s_{max} = 5000$ for  $f_{vv}^{(i)}$ in Eqs.(\ref{conf1}, \ref{conf2},
\ref{conf3}, \ref{conf4}, \ref{conf5}). In addition, the respective error 
deviations for 
$\lambda/L$,$b/L$ and $a/L$ in numerical calculations are $\pm 0.005$,$\pm 
0.0005$ and $\pm 0.0125$.

In the numerical minimization of Eqs.(\ref{conf1}, \ref{conf2},
\ref{conf3}, \ref{conf4}, \ref{conf5}), we take $\ln (4 \Lambda/(e^C \xi )) = 5.57$.
Changing $m \phi_0/\varepsilon_0$ at fixed $\tilde \varepsilon_{dw}$, 
we calculate the minimal energy of each
configuration. We first investigate when these configurations become
energetically favorable in the system. To this end, we check where the
equilibrium energies of the configurations first become negative.

In our analysis, we first identify two regimes of interest: discrete
regime ( $ \varepsilon_{dw} \leq 4 \tilde m^2$ ) and continuum regime  ( $
\varepsilon_{dw} >  4 \tilde m^2$ ). Using our parameters $\tilde
\varepsilon_{dw}$ and $m \phi_0 / \varepsilon_0$, the inequality for
discrete regime  can be expressed as
\begin{equation}
 \tilde \varepsilon_{dw} \leq \frac{\ln (\frac{\Lambda}{\xi})^2}{4 \pi^2}
\left ( \frac{m \phi_0}{ \varepsilon_0 \ln (\frac{\Lambda}{\xi}) } - 1
\right )^2. \label{ineq}
\end{equation}
From
Eq.\ref{ineq}, the minimum value of $m\phi_0/\varepsilon_0$ for
different values of  $\tilde \varepsilon_{dw}$ can be determined. These
values are given in Table.\ref{regime}

\begin{table}[h]
\caption{$\tilde \varepsilon_{dw}$ versus 
$(m\phi_0/\varepsilon_0)_{min}$. The column on the left is input.
\label{regime}}
\begin{center}
\begin{tabular}{|c|c|}\hline
{$\tilde \varepsilon_{dw}$}&{$(m\phi_0/\varepsilon_0)_{min}$}\\
\hline
 0.01 & 6.198 \\ \hline
 0.10 & 7.557 \\ \hline
 1.00 & 11.853 \\ \hline
10.00 & 25.439 \\ \hline
\end{tabular}
\end{center}
\end{table}

When $m\phi_0/\varepsilon_0$ is greater than its minimum value for a
certain $\tilde \varepsilon_{dw}$, the system is then in the discrete
regime. Next, we analyze the proposed cases  for $N=1$ and $N=2$ 
states according to numerical values
given in Table.\ref{regime}. The equilibrium of energies for these cases 
in discrete and continuum regimes are given in Table.\ref{energies}.

\end{multicols}

\begin{table}[h]
\caption{Equilibrum energies for proposed 
configurations. Two columns on the left are input. \label{energies}}
\begin{center}
\begin{tabular}{|c|c|c|c|c|c|c|}\hline
{$\varepsilon_{dw}$}&{$(m\phi_0/\varepsilon_v)_{min}$}&{$\tilde U_{1}$}&
{$\tilde U_{2}$}&{$\tilde U_{3}$}&{$\tilde U_{4}$}&{$\tilde U_{5}$}\\ 
\hline
 0.01 & 5 & -2.58454176 & -2.58455668 & -3.36407195 & -3.36404625 & -3.36404615  \\ \hline
 0.01 & 20 & -65.98296440 & -65.98296500 & -89.20105311 & -89.20030961 &-89.20030943  \\ \hline
 0.1 & 5 & -2.54211623 & -2.54211637 & -3.33057991 & -3.33054828  & -3.32949063 \\ \hline
 0.1 & 20 & -65.87136440 & -65.87136500 & -89.10475311 &-89.10310961 &-89.10310943  \\ \hline
 1  & 5  & -2.16635495 & -2.16637343 & -3.01972350 &-3.01972355 &-3.01972347 \\ \hline
 1  & 20 & -64.78187766 & -64.781878054 & -88.15826391 & -88.15826394 &-88.15826367  \\ \hline
10 & 5 & -0.35644861 & -0.35667498 & -1.21258278 & -1.21232397& -1.21232396\\ \hline 
10 &  25 & -95.00241100 & -95.00247780 & -134.28090345 & -134.27589780& -134.27589769\\ \hline
\end{tabular}
\end{center}
\end{table}

\begin{multicols}{2}

In our numerical calculations,  we find that all proposed configurations are stable both in 
discrete and continuum regimes. This indicates that our method works well 
in both regimes. 
Our numerical calculations also show that the third  configuration wins 
over
the other cases. Nonentheless, this is not enough information for us to 
understand the equilibrium structure, since 
 $\tilde U^{(3)}$  corresponds to two different 
cases with the same structure factor. At this point, we need further 
analysis to determine  
which configuration  is more likely. This can be done  from simple 
physical considerations. Namely, in FSB, the equilibrium structure is 
determined by the competition between vortex-vortex  interaction and vortex - 
magnetization interaction.  
The former favors vortices and antivortices in neighbor domains  
to line up in transverse direction ( perpendicular to magnetic domain wall 
), whereas the latter prefers vortices and antivortices to be shifted so 
that gain in energy is maximized. When vortices are next
to each other on the either side of the magnetic domain wall, the magnetic
fields they produce cancel out each other. From the numerical results, it 
is obvious that vortex-magnetization interaction wins the competion, and 
results in half-way shifting of vortices, if one compares the energies of 
1st and 2nd cases. Then, vortex-magnetization interaction is the dominant 
factor.
By the same token , one can understand what is going on in double vortex 
chain 
configurations. For instance, in the fifth configuration, energy gain due 
to 
vortex-magnetization interaction is diminished, since all the vortices are 
side by side. This explains why equilibrium energy of the fifth 
configuration is higher than those of third and fourth cases.
 In the fourth configuration, the vortices and 
antivortices in the neighbor domains are half-way shifted, so that this 
configuration must be preferred over the one in which they sit side by 
side in the neighbor domains according to the above arguments. However, 
alternative configuration for third case has two chains half-way shifted 
in the neighbor domains instead of one chain as in the fourth case. 
Therefore, one 
might expect gain is even more than that in the fourth configuration.   
Another interesting result is that the system does not favor $N=1$ state 
at all. 
Actually, this does not surprise us since, in the
continuum approximation, we found that the vortex density increases near
the magnetic domain walls. This fact already suggests that the system
favors vortex chains being near the magnetic domain walls rather than a
single chain in the middle of the domain. 


\end{multicols}

\begin{figure}
\centering
\subfigure[1st Case:$L/\Lambda$] 
{
    \label{fig:vstates:a}
    \includegraphics[angle=270,width=7cm]{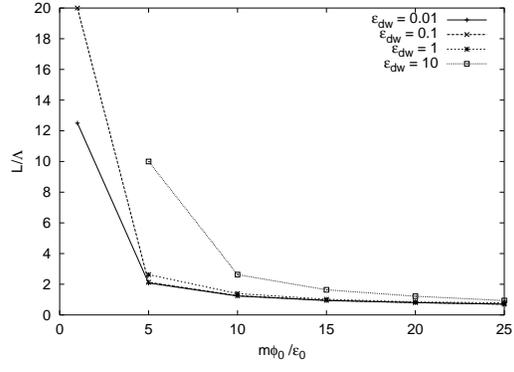}
}
\hspace{1cm}
\subfigure[2nd Case:$L/\Lambda$] 
{
    \label{fig:vstates:b}
    \includegraphics[angle=270,width=7cm]{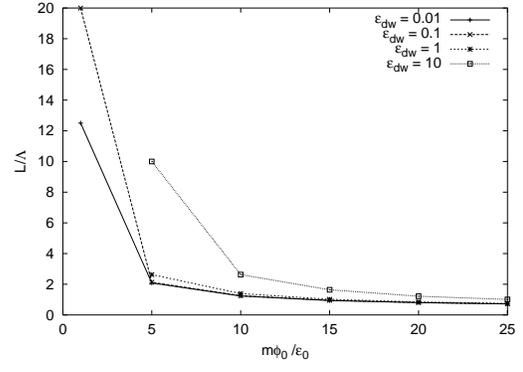}
}
\hspace{1cm}
\subfigure[3rd Case:$L/\Lambda$] 
{
    \label{fig:vstates:c}
    \includegraphics[angle=270,width=7cm]{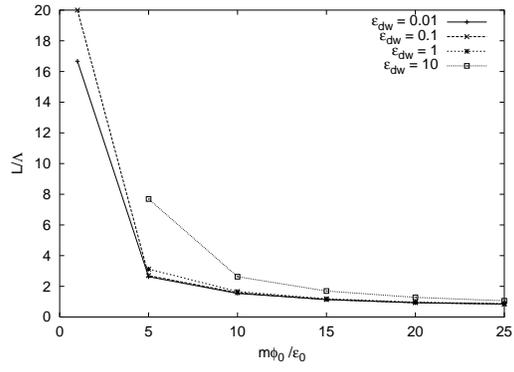}
}
\hspace{1cm}
\subfigure[4th Case:$L/\Lambda$] 
{
    \label{fig:vstates:d}
    \includegraphics[angle=270,width=7cm]{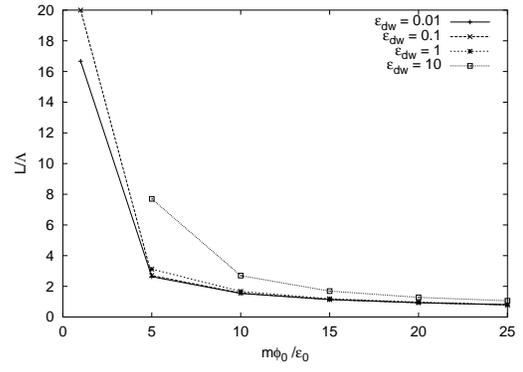}
}
\vspace{1cm}
\subfigure[5th Case:$L/\Lambda$] 
{
    \label{fig:vstates:e}
    \includegraphics[angle=270,width=7cm]{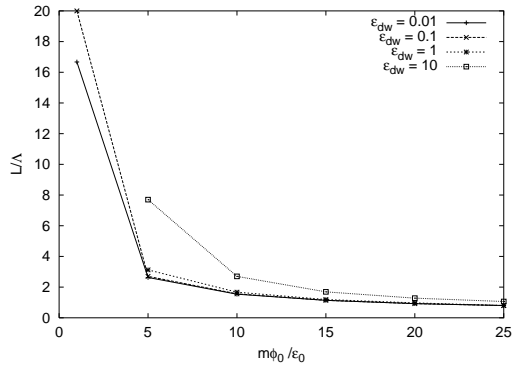}
}
\caption{$L/\Lambda$ versus $m \phi_0/\varepsilon_0$ for $N=1$ and $N=2$ 
states.}
\label{fig:domainsize} 
\end{figure}

\begin{figure}
\centering
\subfigure[1st Case:$b/\Lambda$] 
{
    \label{fig:vstates:a}
    \includegraphics[angle=270,width=7cm]{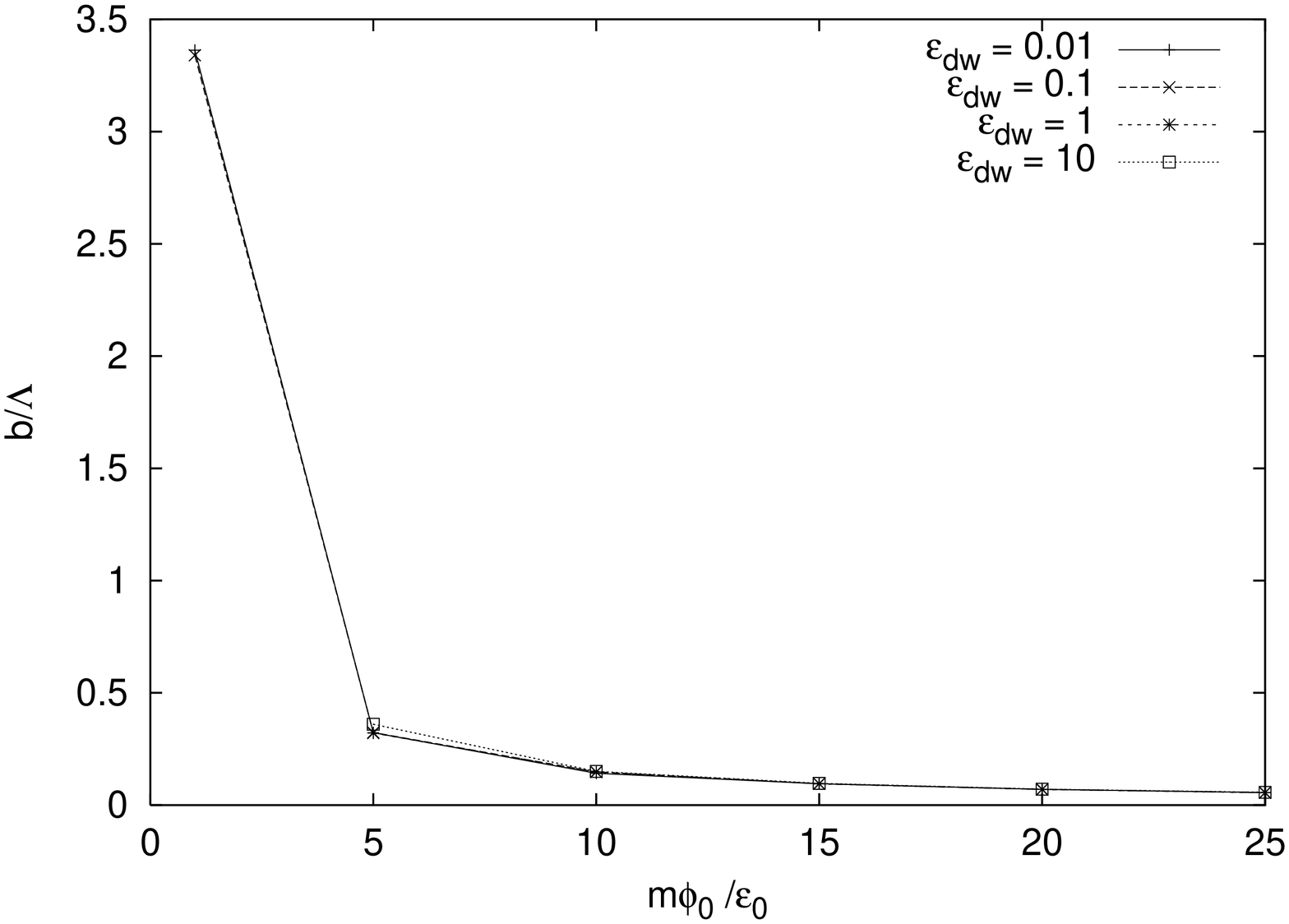}
}
\hspace{1cm}
\subfigure[2nd Case:$b/\Lambda$] 
{
    \label{fig:vstates:b}
    \includegraphics[angle=270,width=7cm]{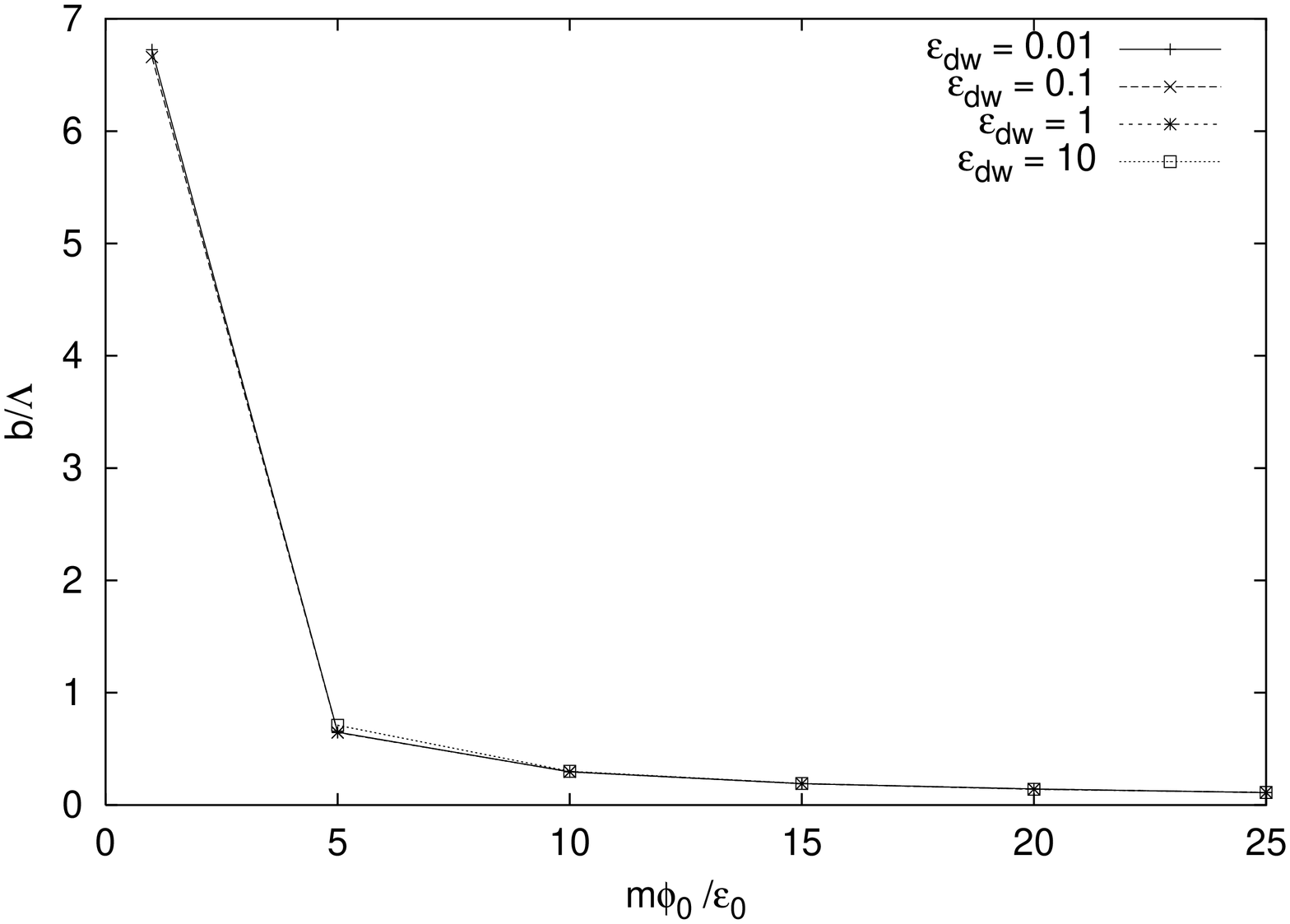}
}
\hspace{1cm}
\subfigure[3rd Case:$b/\Lambda$] 
{
    \label{fig:vstates:c}
    \includegraphics[angle=270,width=7cm]{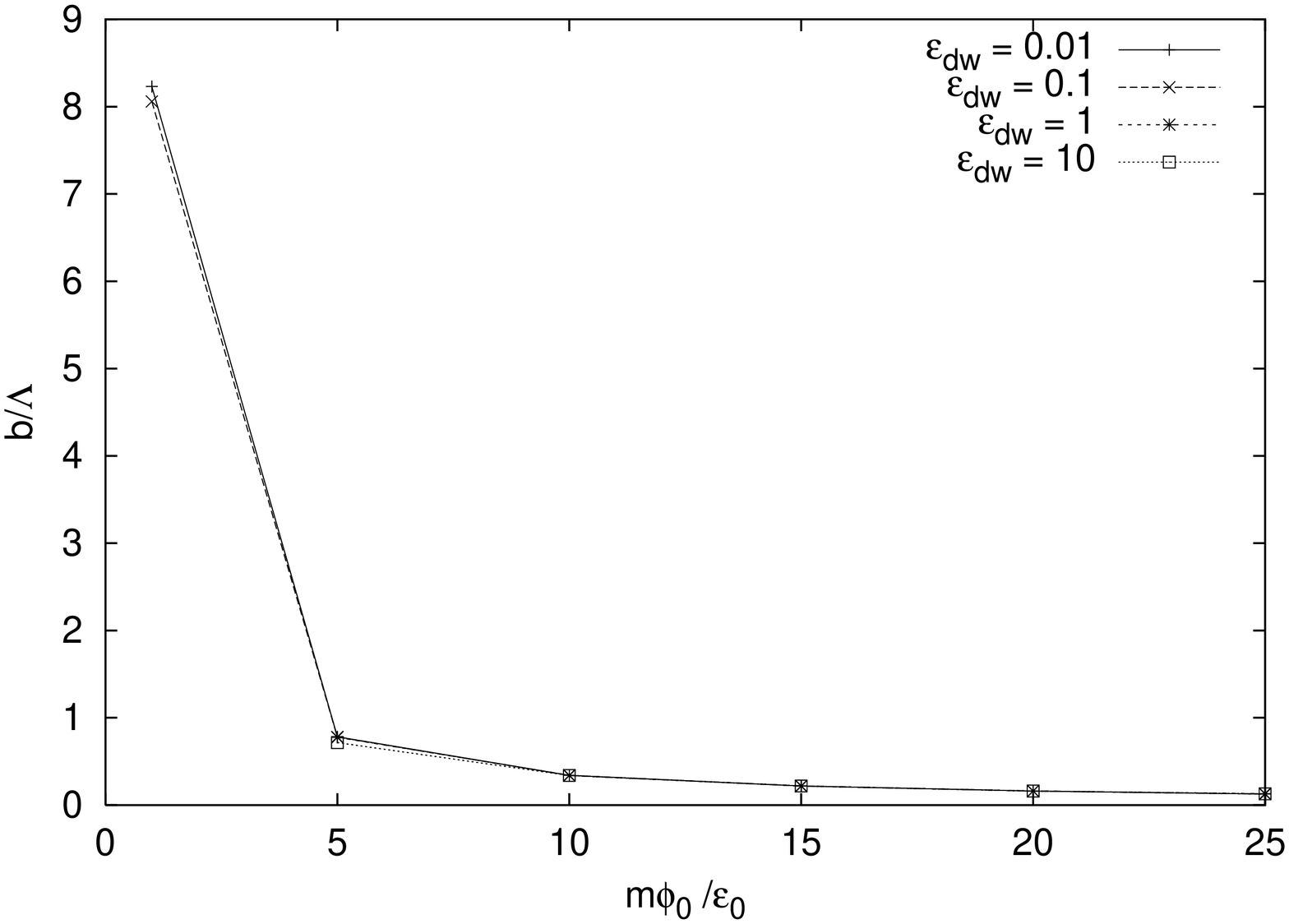}
}
\hspace{1cm}
\subfigure[4th Case:$b/\Lambda$] 
{
    \label{fig:vstates:d}
    \includegraphics[angle=270,width=7cm]{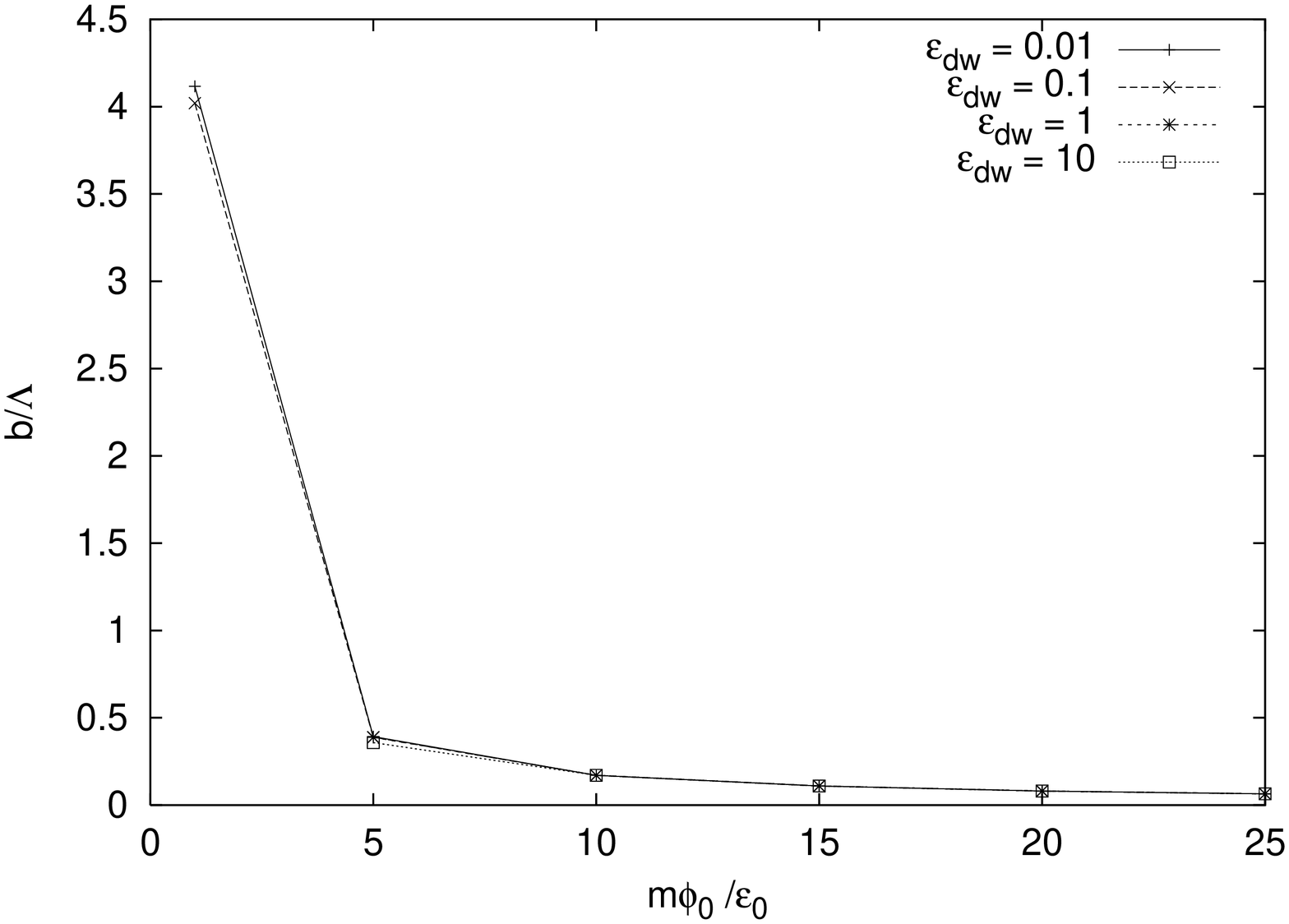}
}
\vspace{1cm}
\subfigure[5th Case:$b/\Lambda$] 
{
    \label{fig:vstates:e}
    \includegraphics[angle=270,width=7cm]{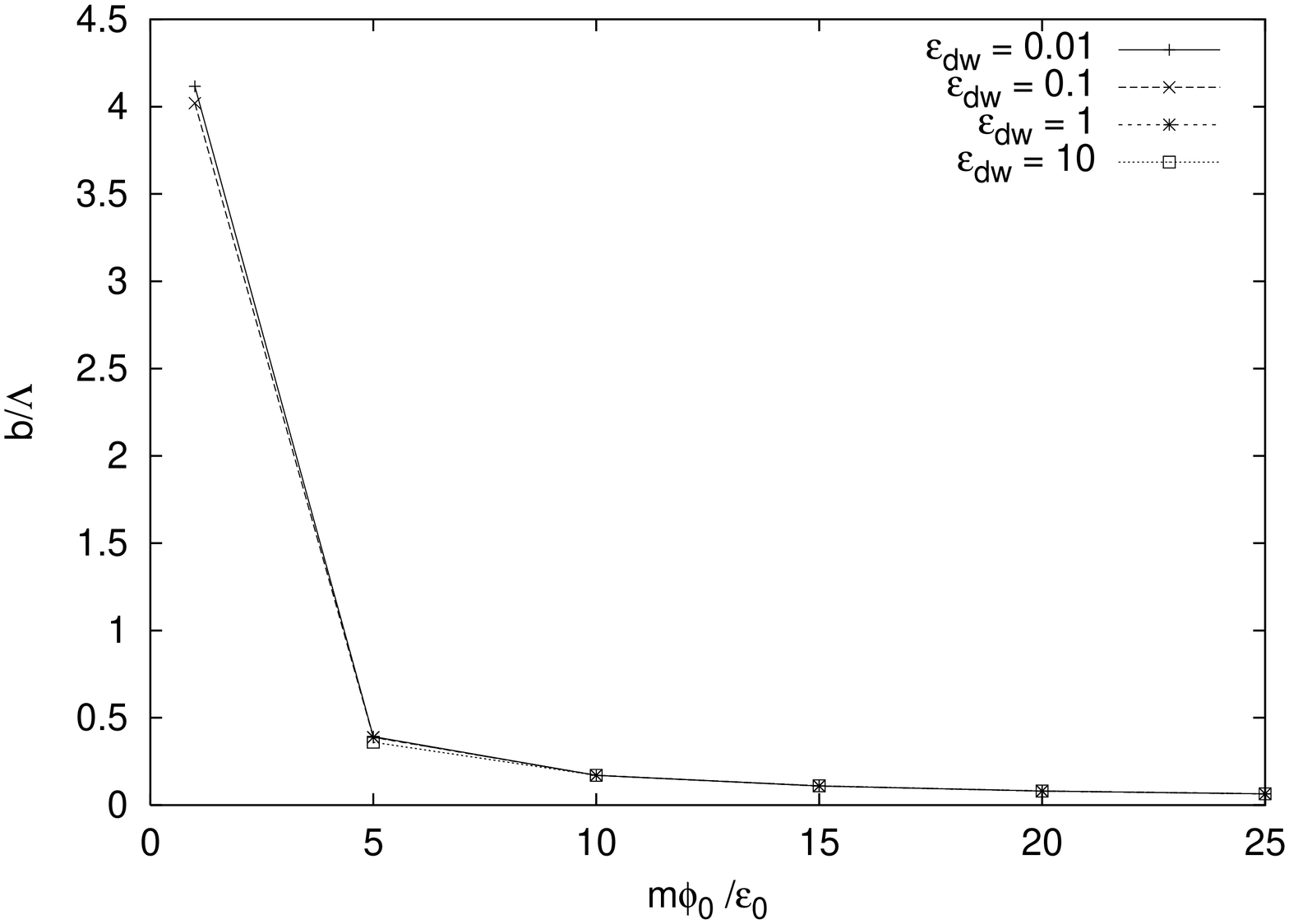}
}
\caption{$b/\Lambda$ versus $m \phi_0/\varepsilon_0$ for $N=1$ and $N=2$ 
states.}
\label{fig:vvdist} 
\end{figure}

\begin{figure}
\centering
\subfigure[3rd Case:$a/\Lambda$] 
{
    \label{fig:vstates:c}
    \includegraphics[angle=270,width=7cm]{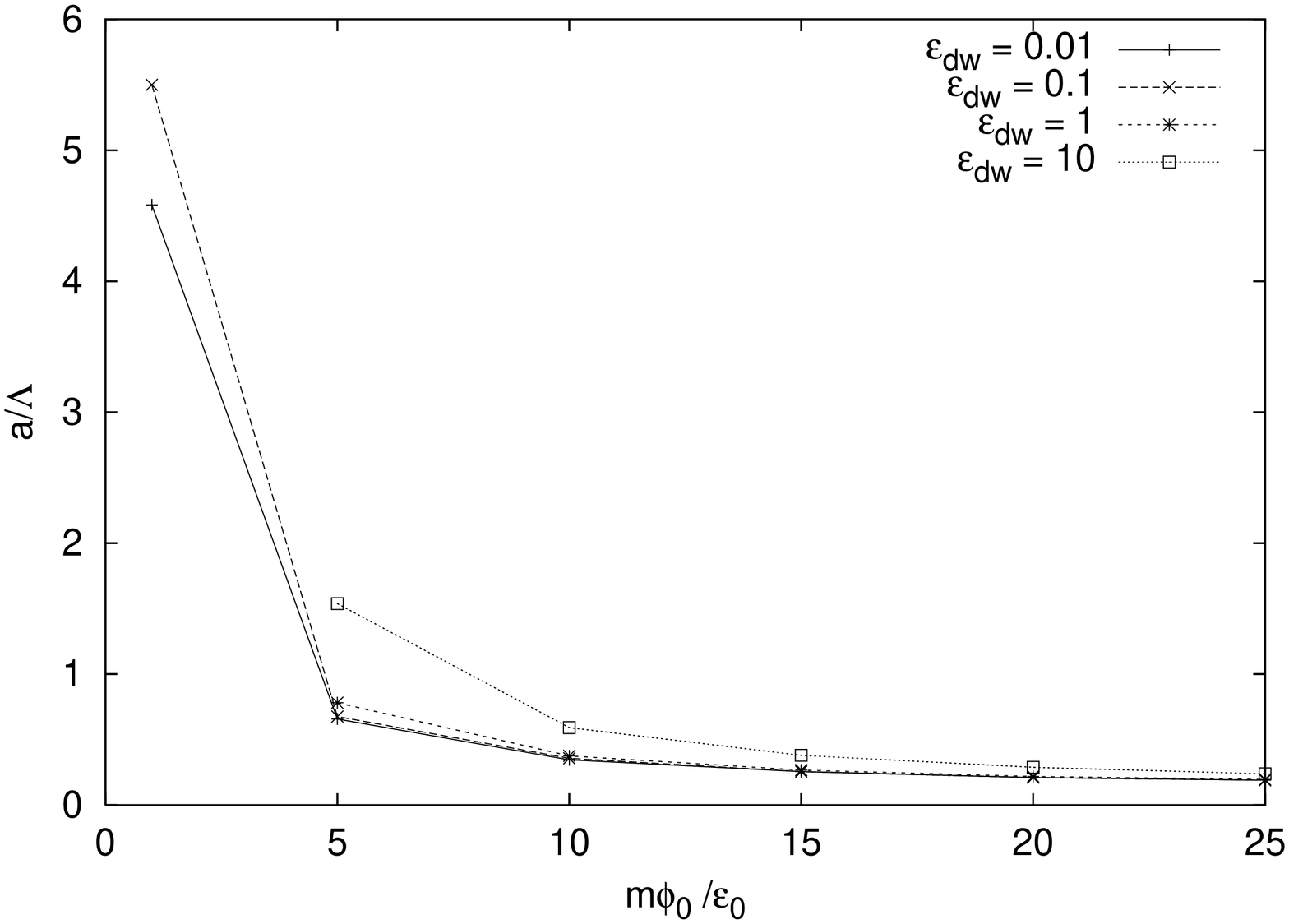}
}
\hspace{1cm}
\subfigure[4th Case:$a/\Lambda$] 
{
    \label{fig:vstates:d}
    \includegraphics[angle=270,width=7cm]{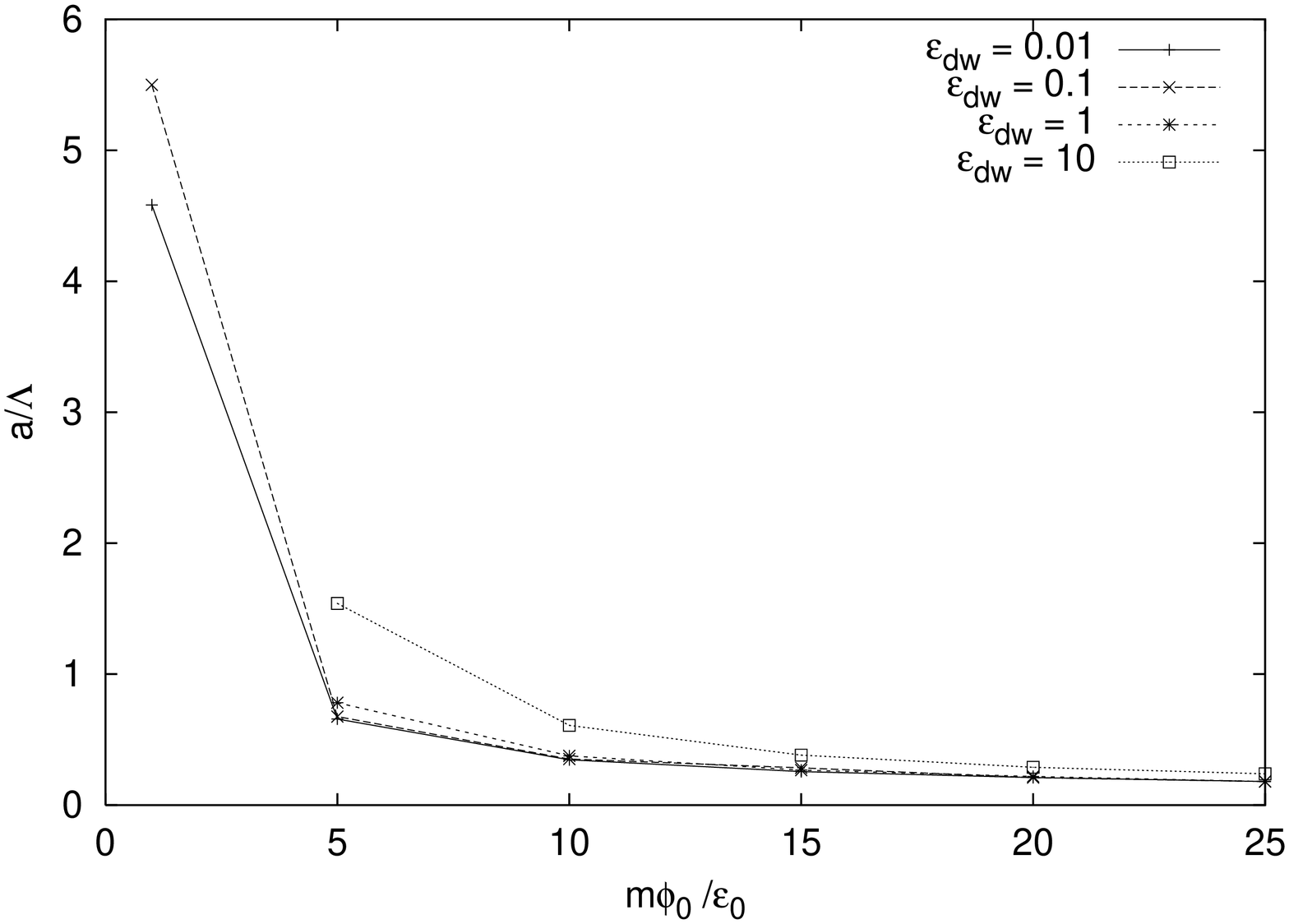}
}
\vspace{1cm}
\subfigure[5th Case:$a/\Lambda$] 
{
    \label{fig:vstates:e}
    \includegraphics[angle=270,width=7cm]{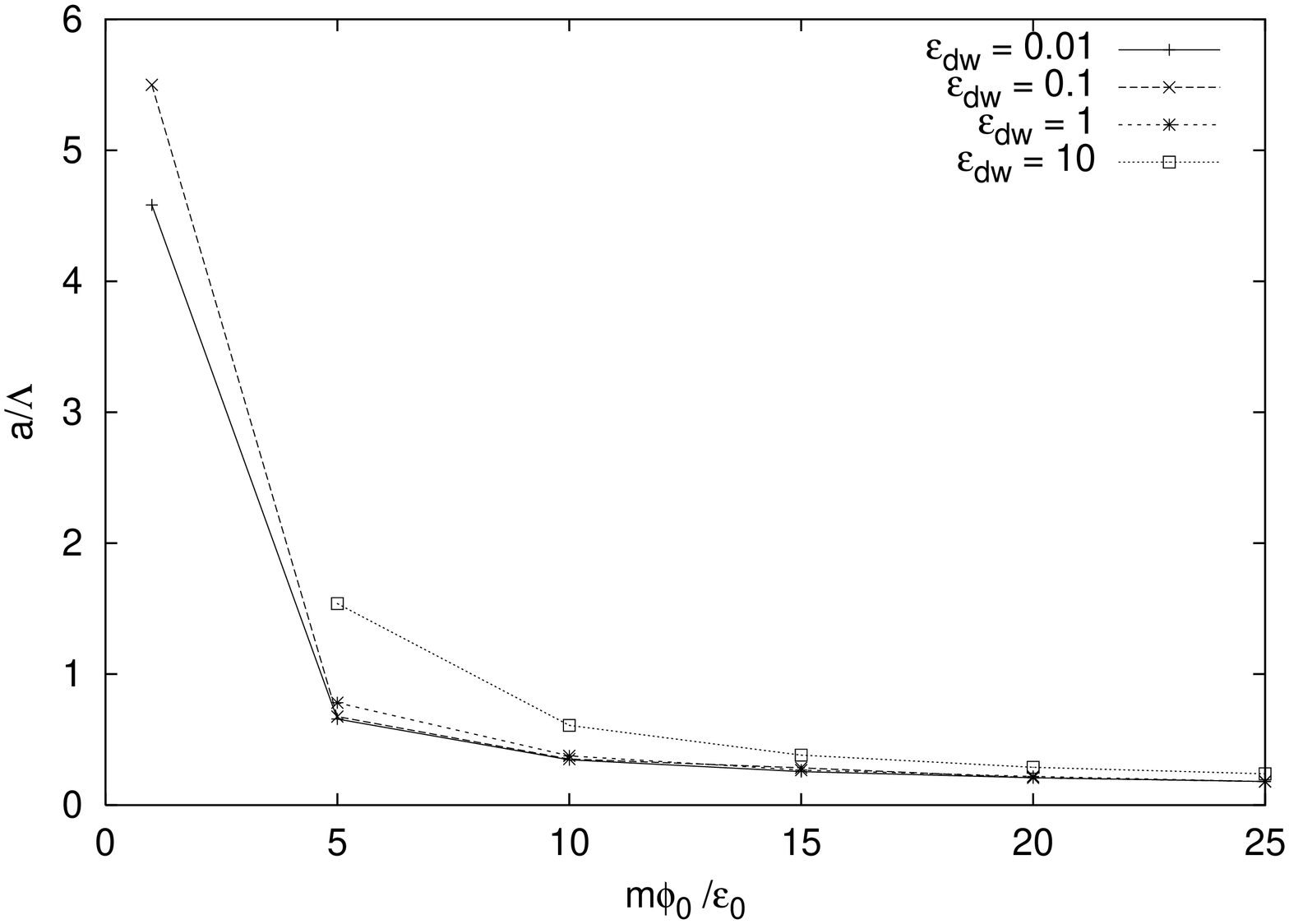}
}
\caption{$a/\Lambda$ versus $m \phi_0 /\varepsilon_0$ for $N=1$ and $N=2$ 
states.}
\label{fig:vmdist} 
\end{figure}

\begin{multicols}{2}

In numerical calculations equilibrium domain size $L/\Lambda$, 
vortex-vortex 
distances  on the same chain $b/\Lambda$ and 
vortex-magnetic domain wall distances $a/\Lambda$ are also calculated. 
These 
results are 
depicted  in Figs.\ref{fig:domainsize},\ref{fig:vvdist},\ref{fig:vmdist}. 
At fixed $\tilde \varepsilon_{dw}$, the further increase of $m \phi_0/\varepsilon_0$
shrinks the domain width, while higher domain wall energy favors larger domain width
at fixed $m \phi_0/\varepsilon_0$, as in usual ferromagnets. That is to say, ferromagnet
favors narrower domains to minimize the demagnetization energy whereas domain wall energy 
makes domains wider. The competition between these two energies determine the domain  
size. Here, the parameter $m \phi_0/\varepsilon_0$ plays the role of 
demagnetization energy.
Domain wall energy does not affect the distance between the 
vortices located on the same chain. However, at larger values of $m \phi_0/\varepsilon_0$
the vortices on the same chain gets closer. This implies that the 
unit  cell area shrinks 
, and consequently vortex density per area increases. Results of 
unit cell areas and vortex densities are discussed with details  in 
the next section.

\section{$N=3$ State}

\begin{figure}[t]
\begin{center}
\includegraphics[angle=0,width=3.5in,totalheight=3.5in]{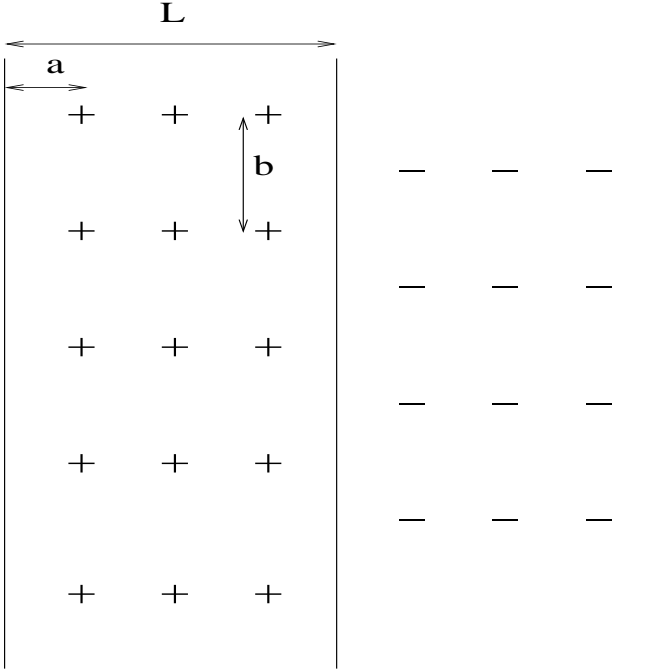}
\caption{The vortex lattice in the first configuration. \label{3cols}}
\end{center}
\end{figure}

In this section, our aim is to understand whether the domain structures 
with different number vortex chains appear spontaneously or not. To answer 
this question, we consider only $N=3$ state and compare its equilibrium 
energy with that of $N=2$ state for various values of magnetic domain wall 
energy 
and magnetization. We first show the derivation of  
energy equation for $N=3$ state. Following the physical arguments in the 
previous section, we 
consider that the equilibrium configuration  is superposition of the 2nd 
case 
and alternative 3rd case (see Fig.\ref{3cols}).

Then, its structure factor
equals to sum of structure factors of those cases, which is given by
\begin{equation}
 F_{\bf G} = (2 i \sin (G_x a)+i (-1)^r ) ( 1 + (-1)^s).
\label{n3sf}
\end{equation}
As seen in second section, self-vortex energy
and  vortex-vortex interaction  depend on $|F_{\bf G}|^2$, whereas $F_{\bf 
G}$ is used in
vortex-magnetization interaction energy.  Then, it is easy 
to see vortex-magnetization energy in this case as the sum
of those of 2nd and 3rd case. We then calculate only the vortex energy 
here.
The square of structure factor is
$|F_{\bf G}|^2 = |F_{\bf G}^{(2)}|^2 + |F_{\bf G}^{(3)}|^2 + (F_{\bf 
G}^{(2)})^*(F_{\bf G}^{(3)})
+ (F_{\bf G}^{(3)})^*(F_{\bf G}^{(2)})$. The first two terms give the 
vortex energy contributions of 2nd and 3rd case. Then,  it is necessary to calculate  
contribution from the cross term only, which  reads
\end{multicols}
\begin{eqnarray}
\tilde U_{cross}  &=& \frac{4 \tilde \Lambda^2}{\tilde b} 
\sum_{r=0}^\infty \sin ( ( 2 r + 1 ) \pi \tilde a) (-1)^r
\left (\frac{ \cot ((2 r + 1 ) \pi \tilde b/4) - 1}{2 r + 1} \right )
+\frac{2\tilde \Lambda^2}{\tilde b} \ln | \cot ( \pi/4 - \pi \tilde a/2 )|
\\ \nonumber
&-& \frac{2 \tilde \Lambda^2}{\pi^2\tilde b^2} \sum_{r,s=-\infty}^{\infty} 
\frac{ \sin ( ( 2 r + 1 ) \pi \tilde a) (-1)^r}
{( 2 r + 1)^2 + 16 s^2/\tilde b^2} \frac{1}{1 + 2 \pi \tilde \Lambda 
\sqrt{( 2 r + 1)^2 + 16 s^2/\tilde b^2}}. \end{eqnarray}
\begin{multicols}{2}
Then, total energy becomes

\end{multicols}
\begin{eqnarray}
\tilde U^{N=3} &=&  \frac{\tilde \Lambda^2}{2\tilde b}
\left(\ln \left( \frac{4 \Lambda}{e^C \tilde \Lambda \xi}\right) + 2f_v^{(2)} ( 
\tilde
\Lambda) - \frac{4 f_{vv}^{(2)} ( \tilde \Lambda, \tilde b ) }{\tilde b
\pi} - \frac{16 m \phi_0}{\varepsilon_0} f_{mv}^{(2)} (\tilde \Lambda
)\right) \\ \nonumber
&+& \frac{\tilde \Lambda^2}{\tilde b}
\left( \ln \left( \frac{4 \Lambda}{e^C \tilde \Lambda \xi}\right)-\ln (\cot(\pi 
\tilde a))
+ 4 f_v^{(3)} ( \tilde \Lambda, \tilde a) - \frac{8 }{\tilde b \pi}
f_{vv}^{(3)} ( \tilde \Lambda, \tilde a, \tilde b ) - \frac{16m
\phi_0}{\varepsilon_0} f_{mv}^{(3)} (\tilde \Lambda,\tilde a )\right) + \tilde 
U_{cross} + \tilde U_{mm} + \tilde \varepsilon_{dw} \tilde \Lambda.
\label{n3en}
\end{eqnarray}

\begin{multicols}{2}
Note that $\tilde U_{mm}$ is described in section II. 
Next, Eq.\ref{n3en} is minimized w.r.t. $\tilde \Lambda$,$\tilde b$ and $\tilde a$ for different 
values of $\tilde \varepsilon_{dw}$ and $m\phi_0/\varepsilon_0$. In 
numerical calculations, the series are calculated as described in the 
previous section. The comporison of equilibrium energies of $N=1$, $N=2$
and $N=3$ states at fixed values of  $\tilde \varepsilon_{dw}$ and $m\phi_0/\varepsilon_0$ are given in 
Table.\ref{comp}.  Note that $N=1$ state represents the second 
configuration in section II, since it is the most energetically favorable 
configuration between two single vortex chain configurations. 

\end{multicols}

\begin{table}[h]
\caption{Equilibrium energies of two chain and three chain states.  Vortex 
density of
unit cell $n$  equals $N_{cell}/S_{el}$, where $N_{cell}= 2 N$ is 
the
number of vortices in a unit cell and $S_{el} = 2 \tilde L \tilde
b$
is equilibrium unit cell area. The two columns on the left are 
input.
\label{comp}}
\begin{center}
\begin{tabular}{|c|c|c|c|c|c|c|c|c|c|c|}\hline
{$\tilde \varepsilon_{dw}$}&{$m\phi_0/\varepsilon_0$}&{$\tilde 
U^{N=1}$}&{$S_{el}$}&{${\it n}$}&{$\tilde 
U^{N=2}$}&{$S_{el}$}&{${\it n}$}&{$\tilde U^{N=3}$}&{$S_{el}$}&{${\it 
n}$}\\
\hline
0.01 & 5 & -2.5846 & 2.70 & 0.74 & -3.3641 & 4.11 & 0.97  & -3.2576 & 6.64 
& 0.90\\ \hline
0.01 & 10 & -13.9742 & 0.73 & 2.74  & -18.5183 &1.04 & 3.82  & -18.8330 & 
1.58 &  3.79 \\ \hline
0.01 & 15 & -35.0799 & 0.36 & 5.56 & -47.0362 & 0.50 & 8.02 &  -48.7311 & 
0.75 & 8.02 \\ \hline
0.01 & 20 & -65.9830 & 0.22 &  8.98 & -89.2011 & 0.30 & 13.24 &  -93.4879 
& 0.44 &  13.49\\ \hline
0.1 & 5 & -2.5421 & 2.75 & 0.73  & -3.3306 & 4.22 & 0.95 &-3.2324 & 6.64 & 
0.90\\ \hline
0.1 & 10 & -13.9015 & 0.74 & 2.71 &-18.4598 & 1.08 &  3.71 &  -18.7780 
&1.58 & 3.79 \\ \hline
0.1 & 15 & -34.9848 & 0.36 &  5.52 & -46.9570 & 0.50 &  8.02 &  -48.6699 & 
0.75 & 8.02 \\ \hline
0.1 & 20 & -65.8714 & 0.23 & 8.74 & -89.1048 & 0.30 & 13.24 & -93.4122 & 
0.44 & 13.49\\ \hline
1 & 5 & -2.1664 & 3.39 & 0.59 &  -3.0197 & 4.84 & 0.83 &  -2.9941 & 
7.13 & 0.84\\ \hline
1 & 10 & -13.2169 & 0.82 & 2.44 & -17.8987 & 1.13 & 3.54 & -18.3421 & 
1.68 & 3.56 
\\ \hline
1 & 15 & -34.0699 & 0.39 & 5.14 & -46.1860 & 0.52 & 7.67 &  -48.0673  & 
0.77 & 7.78\\ \hline
1 & 20 & -67.7819 & 0.24 &  8.40 & -88.1583 & 0.31 & 12.79 & -92.6655 & 
0.45 & 13.18\\ \hline
10 & 5 & -0.3567 & 14.2 &  0.14 &  -1.2126 & 10.99 & 0.36 & -1.4638  & 
14.16 & 0.42\\ \hline
10 & 10 & -8.6308 & 1.58 &  1.27 &  -13.6758 & 1.78 & 2.24 & -14.8269  & 
2.37 &  2.53\\ \hline
10 & 15 & -27.1877 & 0.63 & 3.18 & -39.9099 & 0.74 & 5.41 & -42.8834& 
0.99 & 6.05\\ \hline
10 & 20 & -56.0242 & 0.34 &  5.80 & -80.1062 & 0.41 & 9.66 & -86.0271 & 
0.56 & 10.78 \\ \hline
\end{tabular}
\end{center}
\end{table}

\begin{multicols}{2}

\begin{figure}[t]
\begin{center}
\includegraphics[angle=270,width=3.5in,totalheight=3.5in]{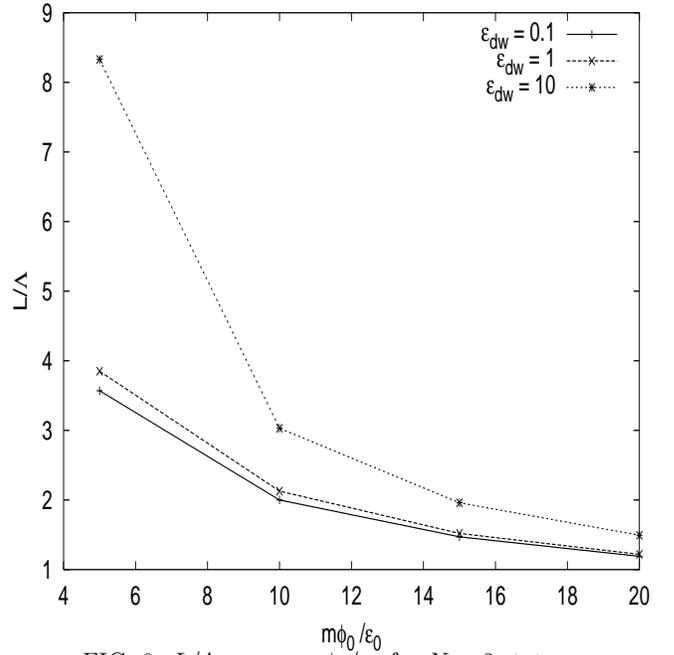}
\caption{$L/\Lambda$ versus $m \phi_0/\varepsilon_0$ for $N=3$ state. 
\label{3coll}}
\end{center}
\end{figure}

\begin{figure}[t]
\begin{center}
\includegraphics[angle=270,width=3.5in,totalheight=3.5in]{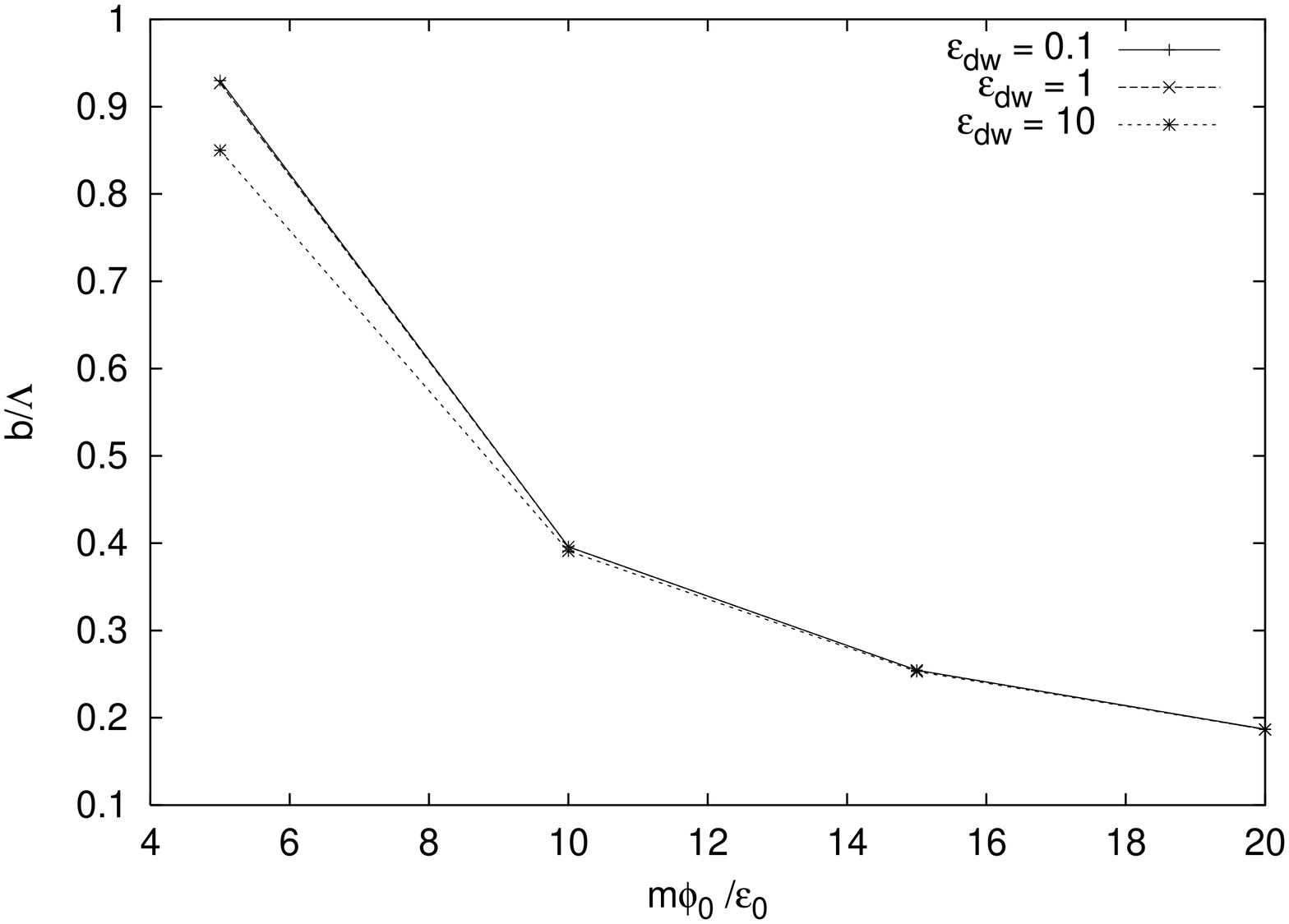}
\caption{$b/\Lambda$ versus $m \phi_0 / \varepsilon_0$  for $N=3$ state. 
\label{3colb}}
\end{center}
\end{figure}

\begin{figure}[t]
\begin{center}
\includegraphics[angle=270,width=3.5in,totalheight=3.5in]{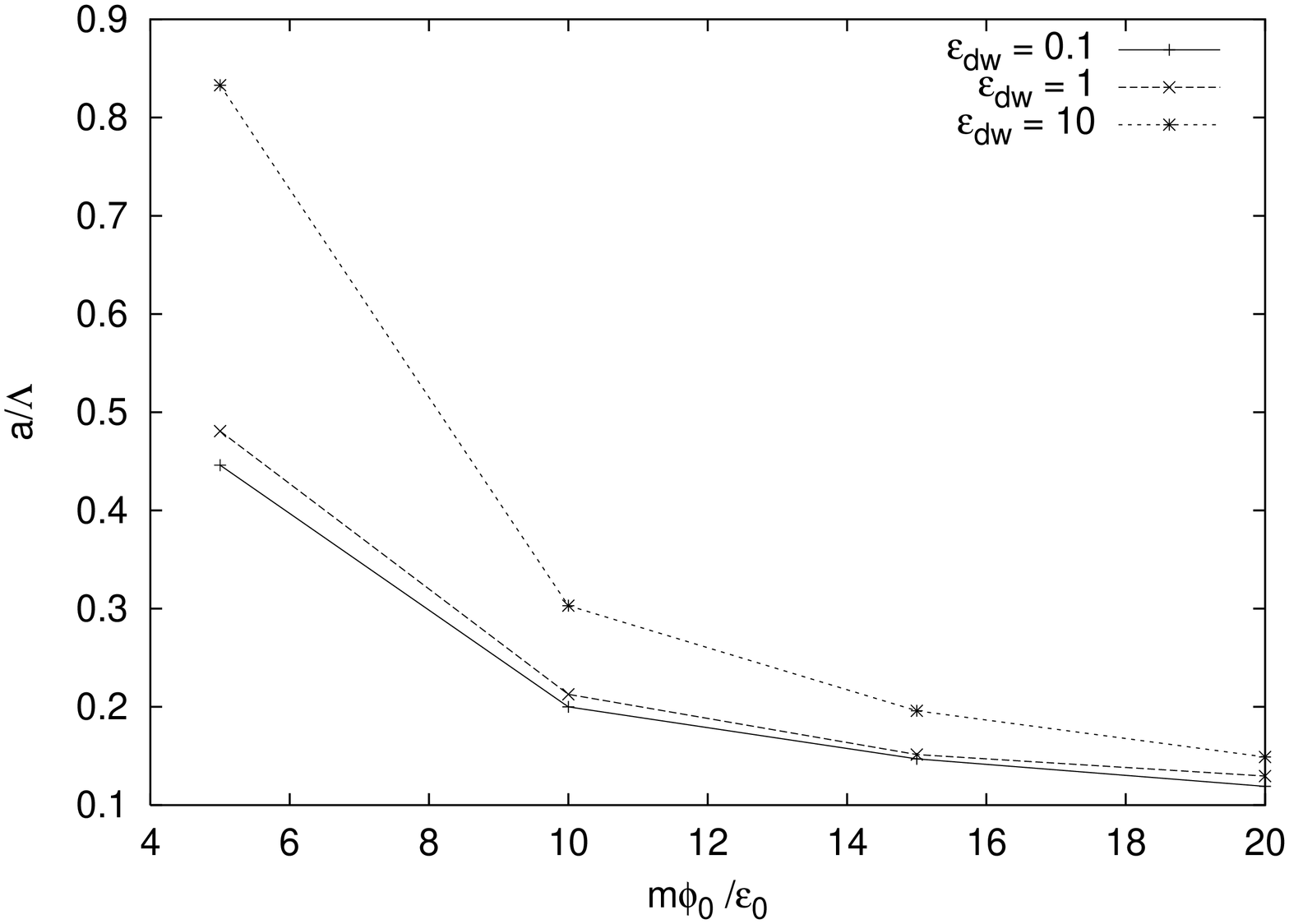}
\caption{$a/\Lambda$ versus $m \phi_0 /\varepsilon_0$  for $N=3$ state. 
\label{3cola}}
\end{center}
\end{figure}

As seen from Table.\ref{comp}, at low magnetization and small domain wall
energies, $N=2$ state is energetically favorable. However, 
$N=3$ state wins over at high values of 
$m\phi_0/\varepsilon_0$ and $\tilde \varepsilon_{dw}$. This picture 
resembles the one we see in the case of magnetic dots on a SC film. In 
that case, the further increase in $m\phi_0/\varepsilon_0$ and the dot's 
size  makes other vortex states more energetically favorable \cite{serkan2}. Here, the 
domain's size is also controlled by domain wall energy. Therefore, the 
larger the $\tilde \varepsilon_{dw}$, the larger the domain size. As a 
result, new chain states might be more energetically favorable.   In $N=3$ state, 
how the equilibrium domain size, vortex-vortex distance on the chain, and 
vortex chain-domain wall distance change according to different values of  
$m\phi_0/\varepsilon_0$ and $\tilde \varepsilon_{dw}$ are shown in 
Figs.\ref{3coll},\ref{3colb},\ref{3cola}. As seen in these figures $L/\Lambda$,$b/\Lambda$
and $a/\Lambda$ follow the same pattern as that of $N=1$ and $N=2$ states. As number of chains per domain 
increases, $L/\Lambda$,$b/\Lambda$
and $a/\Lambda$ increase. 

 In addition
to equilibrium energies, equilibrium unit cell areas and  vortex 
densities of unit cells for  $N=1$,
$N=2$
and $N=3$ states are given in Table.\ref{comp}. At fixed $m
\phi_0/\varepsilon_0$, unit cell area expands with further increase 
of 
$\tilde \varepsilon_{dw}$. However, it shrinks when  $m
\phi_0/\varepsilon_0$ increases. On the other hand, unit cell area 
expands with more more vortex chains at fixed values of $m 
\phi_0/\varepsilon_0$ and $\tilde \varepsilon_{dw}$ except, unit cell 
areal 
of $N=1$ state is larger than those of $N=2$ and $N=3$ states, when $m 
\phi_0/\varepsilon_0 = 5$ and $\tilde \varepsilon_{dw}=10$. We think that 
$N=1$ state in this case is energetically very close to instable region  
in which there 
is monodomain only.   Furthermore, from our 
results, we expect that the most energetically favorable state has the 
highest vortex density.

\section{conclusions}

In this article, we reported our results on the lattices of discrete 
vortices 
in stripe domains in FSB based on  a method based on Maxwell-London 
equations. 
If $\varepsilon_{dw}\leq 4\tilde{m}^2$, the continuum approximation 
becomes invalid. Instead, we considered the discrete lattice of vortices
in which  the vortices were considered to be situated on
chains directed along the stripes. We first analyzed
the vortex configurations up to two vortex chains.
Depending on the magnetization and the
magnetic domain wall energy, the equilibrium energy, the positions of the 
vortices and the
equilbrium
domain
size are calculated.
According to our calculations, in equilibrium, vortices
on the either side of the magnetic domain walls are not side by side on
the
chains; instead,  they are shifted by a half period
along the stripe, while they are side by side in the same domain. We also 
checked whether different vortex chain states can be energetically 
favorable. To do so, we calculated the equilibrium energy for $N=3$ state 
whose equilibrium configuration was assumed to be the superposition of 
those for $N=1$ and $N=2$ states. The comporison of energies implies that  
more vortex chains can emerge spontanously in the larger domains.        

In numerical calculations, we also found that the vortex lattice is stable
for $\varepsilon_{dw} >4 \tilde m^2$. At this point, the domain size is
noticably larger than the effective penetration depth $\Lambda$, so the
continuum approximation is valid. Therefore, we
expect that the domain nucleation starts in the continuum
regime. This problem is left for the future research.  At constant $\tilde \varepsilon_{dw}$, with
increasing $m \phi_0 / \varepsilon_v$, the equilibrium size of the domain
decreases. In addition, the vortices on the chain get closer to each
other. These results agree with those obtained in the continuum
approximation.
As $\varepsilon_{dw}/4\tilde{m}^2$
increases, we expect that new vortex chains develop within the domains. We
leave the detailed analysis of this problem to other publication.

\section{Acknowledgments}
The most of this work was done during my stay at the University of
Minnesota
and was
partially supported by the U.S. Department of Energy, Office of Science,
under Contract No. W-31-109-ENG-38.

\end{multicols}

\appendix

\section*{Calculations of  Series }

\begin{multicols}{2}



In this appendix, the detailed analysis of series is given. First, the
series in the energy calculations of the periodic systems are analyzed;
second, the detailed calculation of the vortex density is shown. The
series we encounter in the energy calculations fall into two categories.
In the  first category, we sum over one variable. The series
in this category are in the form of $\sum_{r=1}^{r_{max}} 1/r$. Employing
the Euler-Maclaurin summation formula \cite{arfken}, the summation is
found with logarithmic accuracy as

\begin{equation} 
\sum_{r=1}^{r_{max}} \frac{1}{r} \approx \ln r_{max} + C.
\label{s1} 
\end{equation}

\noindent where $C \sim 0.577$ is the Euler-Mascheroni constant. If the
summation is performed over only odd integers, we can still transform our
series to (\ref{s1}). Namely,

\begin{eqnarray} 
\sum_{r=0}^{r_{max}} \frac{1}{2 r + 1} &\approx&
\sum_{r=1}^{2 r_{max} +1} \frac{1}{r} - \frac{1}{2} \sum_{r=1}^{r_{max}/2}
\frac{1}{r}, \\ 
&\approx& \ln ( 2 r_{max} + 1 ) + C - \ln
(\frac{r_{max}}{2}) - \frac{C}{2}, \\ &\approx& \frac{1}{2} ( \ln r_{max}
+ C + 2 \ln 2 ). \label{s2new} 
\end{eqnarray}

The other double series of interest here are in the form of

\begin{equation} 
I ( x ) = \sum_{r= -\infty}^{r=\infty} \sum_{s=
-\infty}^{s=\infty} \frac{1}{x^2 r^2 + s^2}, \label{s4new} 
\end{equation}

\noindent where $x$ is an arbitrary constant. Although (\ref{s4new}) is
logarithmically divergent, the sum over one of the variables can be done
easily. To this end, we perform the sum over $s$ first. In doing so, Eq.  
(\ref{s4new}) becomes $(2 \pi/x) \sum_{r=1}^{\infty} \coth (\pi x r)/r$
\cite{series}. This series is logarithmically divergent. In order to get
the logarithmic term , we add and subtract $1/r$. Using the result in
(\ref{s1}), finally we get

\begin{equation} 
I ( x ) \approx \frac{2 \pi}{x} \left[\sum_{r =1}^\infty
\frac{\coth ( \pi x r) - 1}{r} + \ln r_{max} + C\right]. \label{s5}
\end{equation}

\noindent Employing the same techniques, we give the results of the
different versions of Eq. (\ref{s4new}) below:

\end{multicols}


\begin{equation} 
\sum_{r= -\infty}^{r=\infty} \sum_{s= -\infty}^{s=\infty}
\frac{1}{x^2 (2 r + 1)^2 + s^2} \approx
 \frac{2 \pi}{x} \left[\sum_{r=0}^{\infty} \frac{\coth (( 2 r + 1)\pi x ) -1}{2 r + 1} 
+ \frac{\ln r_{max}}{2} + \frac{C}{2}\right], \label{b14}
\end{equation}

\begin{equation} 
\sum_{r= -\infty}^{r=\infty} \sum_{s= -\infty}^{s=\infty}
\frac{1}{x^2 (2 r + 1)^2 + ( 2 s+ 1)^2} \approx
 \frac{ \pi}{x} \left[\sum_{r=0}^{\infty} \frac{\tanh (( 2 r + 1)\frac{\pi
x}{2} ) - 1}{2 r + 1} + \frac{\ln r_{max}}{2} + \frac{C}{2}\right]. \label{b15}
\end{equation}


\begin{multicols}{2}
 
\noindent In (\ref{b14}) and (\ref{b15}), we use $\sum_{s=0}^{\infty}
1/(y^2 + ( 2 s + 1 )^2) = \pi \tanh(\pi y/2)/(4 y)$. In the presence of
$\sin^2 ( (2 r + 1 ) y)$ or $\cos^2 ( (2 r + 1 ) y)$, the series can be
calculated in a similar way, using $\sin^2 ( (2 r + 1 ) y) = (1 - \cos(2 (
2 r + 1)y))/2$ or $\cos^2 ( (2 r + 1 ) y) = (1 + \cos(2 ( 2 r + 1)y))/2$.
For example,

\end{multicols}


\begin{eqnarray} 
\sum_{r= -\infty}^{r=\infty} \sum_{s= -\infty}^{s=\infty}
\frac{\sin^2(( 2 r + 1) y)}{(x^2 (2 r+ 1)^2 + s^2)}
 &=& \frac{2 \pi}{x} \Bigg[\sum_{r=0}^{\infty} \frac{\sin^2 ( (2 r + 1 ) y
)(\coth (( 2 r + 1)\pi x ) - 1)}{2 r + 1}\nonumber \\ 
&+& \frac{\ln r_{max}}{4} - \frac{\ln |\cot(y/2)|}{4}+ \frac{C}{4} \Bigg ], 
\end{eqnarray}

\begin{eqnarray} 
\sum_{r= -\infty}^{r=\infty} \sum_{s= -\infty}^{s=\infty}
\frac{\cos^2(( 2 r + 1) y)}{(x^2 (2 r+ 1)^2 + s^2)}
 &=& \frac{2 \pi}{x} \Bigg [\sum_{r=0}^{\infty} \frac{\sin^2 ( (2 r + 1 ) y
)(\coth (( 2 r + 1)\pi x ) - 1)}{2 r + 1}\nonumber \\ 
&+& \frac{\ln
r_{max}}{4} + \frac{\ln |\cot(y/2)|}{4}+ \frac{C}{4} \Bigg ]. 
\end{eqnarray}

\noindent We use 
\begin{equation} 
\sum_{r=0}^\infty \frac{\cos((2 r +1
)\theta)}{2 r +1 } = \frac{\ln | \cot (\theta/2)|}{2}. \label{coth}
\end{equation} 



\begin{multicols}{2}

\end{multicols}

\end{document}